\documentclass[12pt]{iopart}
\eqnobysec
\usepackage{iopams} 
\usepackage{amsthm} 
\usepackage{graphicx}
\usepackage{cite}

\newcommand{\pebra}{\ensuremath{\langle ^+ \! \hskip0.02cm E|}}
\newcommand{\nebra}{\ensuremath{\langle ^- \! \hskip0.02cm E|}}

\newcommand{\Sw}{{\cal S}(\mathbb R ^+ \! \frac{\ }{\ } \{ a,b \} )}
\newcommand{\Swt}{{\cal S}^{\times}(\mathbb R ^+ \!\frac{\ }{\ } \{ a,b \} )}
\newcommand{\Swp}{{\cal S}^{\prime}(\mathbb R \frac{\ }{\ } \{ a,b \} )}
\newcommand{\rhsSwt}{\Sw \subset L^2(\mathbb R ^+, \rmd r) \subset \Swt}

\newcommand{\Swhp}{\widehat{{\cal S}}_+(\mathbb R ^+ \! \frac{\ }{\ }\{ a,b \})}\newcommand{\Swhpt}{\widehat{{\cal S}}_+^{\times}(\mathbb R ^+ \! \frac{\ }{\ }\{ a,b \})}
\newcommand{\Swhn}{\widehat{{\cal S}}_-(\mathbb R ^+ \! \frac{\ }{\ }\{ a,b \})}\newcommand{\Swhnt}{\widehat{{\cal S}}_-^{\times}(\mathbb R ^+ \! \frac{\ }{\ }\{ a,b \})}

\begin{document}

\def\llra{\relbar\joinrel\longrightarrow}              
\def\mapright#1{\smash{\mathop{\llra}\limits_{#1}}}    
\def\mapup#1{\smash{\mathop{\llra}\limits^{#1}}}     
\def\mapupdown#1#2{\smash{\mathop{\llra}\limits^{#1}_{#2}}} 

\catcode`\@=11

\def\BF#1{{\bf {#1}}}
\def\NEG#1{{\rlap/#1}}

\def\Let@{\relax\iffalse{\fi\let\\=\cr\iffalse}\fi}
\def\vspace@{\def\vspace##1{\crcr\noalign{\vskip##1\relax}}}
\def\multilimits@{\bgroup\vspace@\Let@
 \baselineskip\fontdimen10 \scriptfont\tw@
 \advance\baselineskip\fontdimen12 \scriptfont\tw@
 \lineskip\thr@@\fontdimen8 \scriptfont\thr@@
 \lineskiplimit\lineskip
 \vbox\bgroup\ialign\bgroup\hfil$\m@th\scriptstyle{##}$\hfil\crcr}
\def\Sb{_\multilimits@}
\def\endSb{\crcr\egroup\egroup\egroup}
\def\Sp{^\multilimits@}
\let\endSp\endSb


\title[The RHS approach to the Lippmann-Schwinger equation]{The rigged 
Hilbert space approach to the Lippmann-Schwinger equation. Part I}

\author{Rafael de la Madrid}
\address{Department of Physics, University of 
California at San Diego, La Jolla, CA 92093 \\
E-mail: {\texttt{rafa@physics.ucsd.edu}}}

\begin{abstract}
We exemplify the way the rigged Hilbert space deals with the 
Lippmann-Schwinger equation by way of the spherical shell potential. We
explicitly construct the Lippmann-Schwinger bras and kets along with
their energy representation, their time evolution and the rigged 
Hilbert spaces to which they belong. It will be concluded that the natural 
setting for the solutions of the Lippmann-Schwinger equation---and therefore 
for scattering theory---is the rigged Hilbert space rather than just the 
Hilbert space.
\end{abstract}




\section{Introduction}
\label{sec:introduction}

Scattering experiments play a central role in our understanding of the
microscopic world. For example, Compton scattering of x rays by electrons is 
held as experimental evidence for the particle nature of the photon; much of 
what has been learned about the structure of the nucleus, indeed even its 
discovery, was the results of scattering experiments; the analysis of 
scattering has yielded most of our present knowledge of elementary particle 
physics~\cite{PDT}. Scattering theory was developed in order to understand 
such experiments.

The Lippmann-Schwinger equation is one of the cornerstones of scattering
theory. It was introduced by Lippmann and Schwinger in 1950~\cite{LS}, and 
it has become a commonplace in scattering 
theory~\cite{GMG,GOLDBERGER,NEWTON,TAYLOR}. It is written as
\begin{equation}
       |E ^{\pm}\rangle =|E\rangle +
       \frac{1}{E-H_0\pm \rmi \varepsilon}V|E^{\pm}\rangle \, ,
       \label{LSeq1}
\end{equation}
where $|E^{\pm}\rangle$ represent the ``in'' and ``out'' Lippmann-Schwinger 
kets, $|E\rangle$ represents an eigenket of the free Hamiltonian $H_0$,
\begin{equation}
      H_0|E\rangle =E|E\rangle \, ,
      \label{tisequa0}
\end{equation}
and $V$ is the potential. The Lippmann-Schwinger kets are, in particular,
eigenvectors of $H$:
\begin{equation}
      H|E^{\pm}\rangle =E|E ^{\pm}\rangle \, .
      \label{tisequa}
\end{equation}
To the kets $|E^{\pm}\rangle$, there correspond the bras $\langle ^{\pm}E|$, 
which satisfy
\begin{equation}
        \langle ^{\pm}E| = 
        \langle E| +\langle ^{\pm}E|V\frac{1}{E-H_0\mp \rmi \varepsilon} \, .
       \label{LSeqbraspm}
\end{equation}
The bras $\langle ^{\pm}E|$ are left eigenvectors of $H$,
\begin{equation}
        \langle ^{\pm}E|H = E \langle ^{\pm}E|  \, ,
       \label{LSeibenbraeq}
\end{equation}
and the bras $\langle E|$ are left eigenvectors of $H_0$,
\begin{equation}
        \langle E|H = E \langle E|  \, .
       \label{LSeibenbraeq0}
\end{equation}

In spite of its wide acceptance, the Lippmann-Schwinger equation still lacks
a proper mathematical setting. There are mainly two problems. First, the 
bras $\langle ^{\pm}E|$ and the kets $|E^{\pm}\rangle$ are not square 
integrable (i.e., they lie outside the Hilbert space), and therefore they 
must be treated as distributions. And second, in resonance theory the bras 
$\langle ^{\pm}E|$ and the kets $|E^{\pm}\rangle$ need to be analytically
continued into the complex 
plane~\cite{MONDRA,GADELLA84,BG,GADELLA-ORDONEZ,BRTK,CAGADA,GAGO,DIS}. Both 
problems can be handled by means of the rigged Hilbert space 
(RHS). In this paper we shall focus on the first problem, namely on how to 
accommodate the Lippmann-Schwinger bras and kets for real, positive energies 
within the RHS setting. The second problem will be addressed in a sequel to
this paper.

The present paper is the continuation of Refs.~\cite{JPA02,FP02,IJTP03,JPA04},
where we showed how the RHS deals with continuous spectrum. As we shall 
see, the methods of Refs.~\cite{JPA02,FP02,IJTP03,JPA04} can be 
straightforwardly applied to the Lippmann-Schwinger equation.

We recall that a rigged Hilbert space is a triad of spaces (see
Ref.~\cite{EJP05} for a pedagogical introduction to the RHS)
\begin{equation}
       {\mathbf \Phi} \subset {\cal H} \subset {\mathbf \Phi}^{\times}
      \label{RHStIntro}
\end{equation}
such that $\cal H$ is a Hilbert space, $\mathbf \Phi$ is a dense
subspace of $\cal H$, and $\mathbf \Phi ^{\times}$ is the space of
antilinear functionals over $\mathbf \Phi$. The space $\mathbf \Phi ^{\times}$
is called the antidual space of $\mathbf \Phi$. The space $\mathbf \Phi$ is 
the largest subspace of the Hilbert space such that first, it remains 
invariant under the action of the Hamiltonian, and second, the action of 
the Lippmann-Schwinger bras and kets is well defined on its 
elements. Associated with the RHS~(\ref{RHStIntro}), there is always another 
RHS,
\begin{equation}
       {\mathbf \Phi} \subset {\cal H} \subset {\mathbf \Phi}^{\prime} \, ,
      \label{RHSpIntro}
\end{equation}
where ${\mathbf \Phi}^{\prime}$ is called the dual space of ${\mathbf \Phi}$
and contains the linear functionals over $\mathbf \Phi$. As we shall show,
the Lippmann-Schwinger bras $\langle ^{\pm}E|$ belong to 
${\mathbf \Phi}^{\prime}$, whereas the Lippmann-Schwinger kets 
$|E^{\pm} \rangle$ belong to ${\mathbf \Phi}^{\times}$.

It should be emphasized that our RHS does not entail an extension of the 
physical principles of scattering theory. Our RHS simply enables us to 
extract and process information contained in the Lippmann-Schwinger bras 
and kets. 

There have been previous attempts at describing the Lippmann-Schwinger 
equation within the RHS setting in 
Refs.~\cite{GADELLA84,BG,GADELLA-ORDONEZ,BRTK,CAGADA,GAGO,DIS}. Our results
will be different from those 
of~\cite{GADELLA84,BG,GADELLA-ORDONEZ,BRTK,CAGADA,GAGO,DIS} in the following
sense. First, Refs.~\cite{GADELLA84,BG,GADELLA-ORDONEZ,BRTK,CAGADA,GAGO,DIS}
use formal manipulations based on the Hardy-space assumption, whereas
our approach is based on an explicitly solvable example; what is more, we 
do not make any assumption concerning Hardy functions, because Hardy functions 
seem unrelated to the solutions of the Lippmann-Schwinger equation. Second,
Refs.~\cite{GADELLA84,BG,GADELLA-ORDONEZ,BRTK,CAGADA,GAGO,DIS} use two
different RHSs (of Hardy functions), whereas we shall use only one RHS. Third,
the Hardy-function assumption of
Refs.~\cite{GADELLA84,BG,GADELLA-ORDONEZ,BRTK,CAGADA,GAGO,DIS} implies that
the time evolution of the Lippmann-Schwinger bras and kets associated with
real energies is given by a semigroup, whereas in this paper the time evolution
of the Lippmann-Schwinger bras and kets is given by a group. And fourth,
using Hardy functions as in
Refs.~\cite{GADELLA84,BG,GADELLA-ORDONEZ,BRTK,CAGADA,GAGO,DIS} entails
an extension of the physical principles of scattering theory, whereas in
this paper we shall remain within standard quantum scattering theory.

The paper is organized as follows: In Sec.~\ref{sec:AROSTH}, we recall some
of the basics of scattering theory. In Sec.~\ref{sec:LSequation}, we 
obtain the Lippmann-Schwinger eigenfunctions in the radial, position 
representation for the spherical shell potential and for zero angular 
momentum. In Sec.~\ref{sec:dospresolvent}, we recall the expressions for the 
domain, the spectrum and the Green function of the Hamiltonian. In 
Sec.~\ref{sec:DIdecomop}, we obtain the unitary operators $U_{\pm}$, 
the eigenfunction expansions and the direct integral 
decompositions generated by the Lippmann-Schwinger eigenfunctions. In 
Sec.~\ref{sec:consLSbkin-out}, we construct the Lippmann-Schwinger bras and 
kets from the corresponding eigenfunctions, and we obtain their most relevant 
properties. In Sec.~\ref{sec:timevoLSbk}, we calculate the time evolution
of the Lippmann-Schwinger bras and kets. In Sec.~\ref{sec:otherreusst}, we 
justify some standard results of scattering theory within the RHS setting. In 
Sec.~\ref{sec:conclusions}, we state our conclusions.

\section{A reminder of scattering theory}
\label{sec:AROSTH}

\begin{figure}[ht]
\includegraphics[width=15cm]{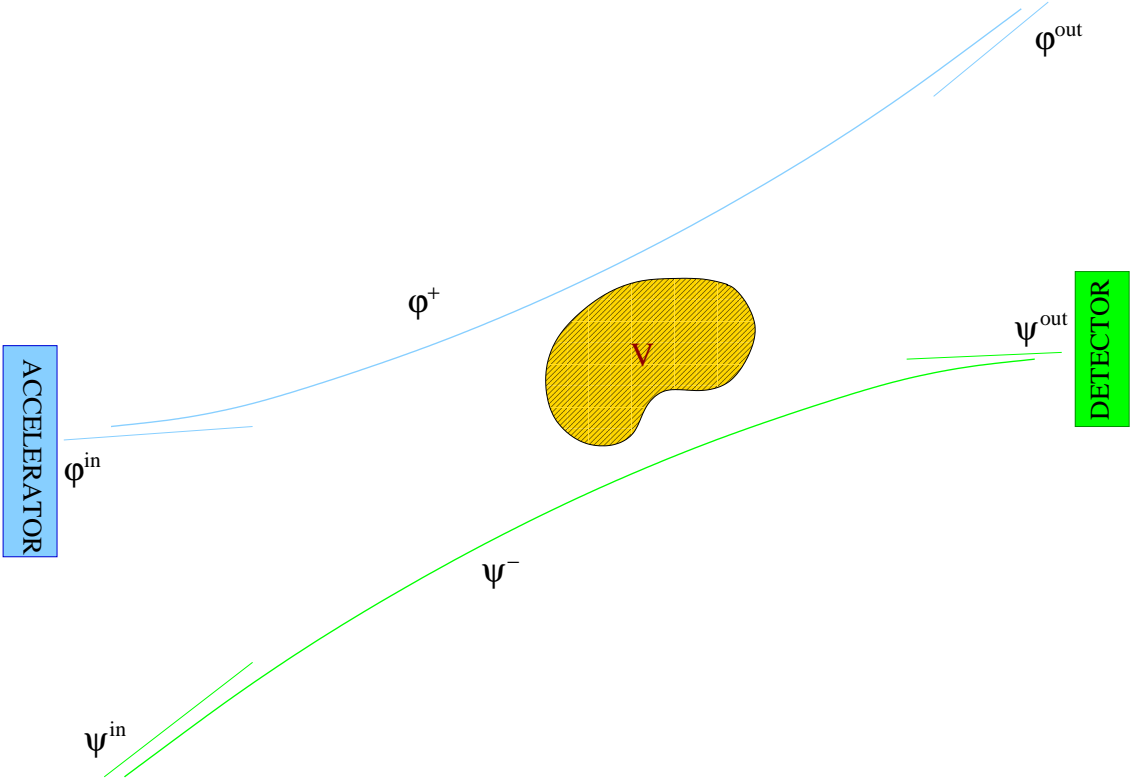}
\vskip0.2cm
\caption{Schematic, pictorial representation of a scattering process.}
\label{fig:scattering}
\end{figure}

A scattering process can be pictorially summarized as in 
Fig.~\ref{fig:scattering}. Loosely 
speaking, we send a beam of prepared initial ``in'' states 
$\varphi ^{\rm in}$ toward the potential. After the
collision takes place, $\varphi ^{\rm in}$ becomes 
$\varphi ^{\rm out}$. We then measure the probability to find 
a final ``out'' state $\psi ^{\rm out}$. The amplitude of this probability is 
given by
\begin{equation}
      (\psi ^{\rm out},\varphi ^{\rm out})=
      (\psi ^{\rm out},S\varphi ^{\rm in}) \, ,
      \label{iprobaoutin}  
\end{equation}
where $S$ is the $S$~matrix. The state $\varphi ^{\rm in}$ is 
determined by {\it initial} conditions through a preparation procedure 
(e.g., an accelerator). The state $\varphi ^{\rm out}=S\varphi ^{\rm in}$ is 
determined by those initial conditions through $\varphi ^{\rm in}$ and by
the scattering process through the $S$ matrix. The state 
$\psi ^{\rm out}$ is determined by {\it final} conditions through a 
registration procedure (e.g., a detector). 

The initial ``in'' state $\varphi ^{\rm in}$ 
and the final ``out'' state $\psi ^{\rm out}$ are asymptotic forms of the 
so-called ``in'' state $\varphi ^+$ and ``out'' state $\psi ^-$ in the remote 
past and in the distant future, respectively. In terms of these, the 
probability amplitude (\ref{iprobaoutin}) reads
\begin{equation}
      (\psi ^{\rm out},\varphi ^{\rm out})=(\psi ^{-},\varphi ^{+}) \, .
      \label{iprobaout+-}  
\end{equation}
The asymptotic states $\varphi ^{\rm in}$ and $\psi ^{\rm out}$ are related 
to the ``exact'' states $\varphi ^+$ and $\psi ^-$ by the M{\o}ller operators,
\numparts
\begin{eqnarray}
       &&\Omega _+\varphi ^{\rm in}=\varphi ^+ \, , \\
       &&\Omega _-\psi ^{\rm out}=\psi ^- \, .
\end{eqnarray}
\endnumparts

It is important to keep in mind that the states $\varphi ^{\rm in}$, 
$\varphi ^{\rm out}$ and $\psi ^{\rm out}$ (or $\varphi ^+$ and $\psi ^-$) are
not the central object of quantum scattering. The central object
of quantum scattering is the probability 
amplitude~(\ref{iprobaoutin})-(\ref{iprobaout+-}), which measures the overlap 
between the ``in'' (i.e., prepared) and the ``out'' (i.e., detected) 
states. This is in distinction to classical scattering, where the states
$\varphi ^{\rm in}$, $\varphi ^+$ and $\varphi ^{\rm out}$ do actually 
represent a physical wave, and where we actually are able 
to observe $\varphi ^{\rm out}$ rather than the overlap of 
$\varphi ^{\rm out}$ with some $\psi ^{\rm out}$.

It is customary to split up the (total) Hamiltonian $H$ into the free 
Hamiltonian $H_0$ and the potential $V$,
\begin{equation}
       H=H_0+V \, .
\end{equation}
The potential $V$ is interpreted as the interaction between the components of 
the initial prepared states, for instance, the interaction between the 
in-going beam and the target. The states $\varphi ^{\rm in}$ and 
$\psi ^{\rm out}$ evolve under the influence of the free Hamiltonian $H_0$, 
whereas the states $\varphi ^{+}$ and $\psi ^{-}$ evolve under the influence 
of the (total) Hamiltonian $H$. 

The dynamics of a scattering system is therefore governed by the Schr\"odinger
equation subject to boundary conditions that specify what is ``in'' and what 
is ``out.''  The Lippmann-Schwinger equation~(\ref{LSeq1}) for the ``in''
$|E^{+}\rangle$ and ``out'' $|E^{-}\rangle$ kets has those ``in'' and ``out'' 
boundary conditions built into the $\pm \rmi \varepsilon$, since 
Eq.~(\ref{LSeq1}) is equivalent to the time-independent Schr\"odinger 
equation~(\ref{tisequa}) subject to those ``in'' ($+\rmi \varepsilon$) and 
``out'' ($-\rmi \varepsilon$) boundary conditions. In the position 
representation, the $+\rmi \varepsilon$ prescription yields the following 
asymptotic behavior:
\begin{equation}
      \langle {\bf x}|E^+\rangle \, \mapupdown{}{r\to \infty} \
      \rme^{\rmi kz}+f(\theta , \phi) \, \frac{\rme^{\rmi kr}}{r} \, ,
       \label{asympbepk}
\end{equation}
where ${\bf x}\equiv (x,y,z) \equiv (r,\theta , \phi)$ are the position
coordinates, $k$ is the wave number and $f(\theta ,\phi )$ 
is the so-called scattering amplitude. Thus, far away from the potential
region, $\langle {\bf x}|E^+\rangle$ is 
a linear combination of a plane wave (which originates from the free part 
in Eq.~(\ref{LSeq1})) and an outgoing spherical wave multiplied by the
scattering amplitude (which originate from the second term on the
right-hand side of Eq.~(\ref{LSeq1}), and which account for the
effect of the potential on the incoming beam). The $-\rmi \varepsilon$ 
prescription leads to the following asymptotic behavior:
\begin{equation}
      \langle {\bf x}|E^-\rangle \, \mapupdown{}{r \to \infty} \
      \rme ^{\rmi kz} + \overline{f(\theta ,\phi )} \, 
           \frac{\rme ^{-\rmi kr}}{r} \, ;
      \label{asympbemk}
\end{equation}
thus, far away from the potential region, $\langle {\bf x}|E^-\rangle$ is 
a combination of a plane wave and an incoming spherical wave multiplied by 
the complex conjugate of $f(\theta ,\phi )$.

Formally, the ``in'' state $\varphi ^+$ and the ``out'' state $\psi ^-$ can 
be expanded in terms of the Lippmann-Schwinger bras and kets as follows:
\begin{eqnarray}
       &&\varphi ^+=\int_{0}^{\infty} \rmd E \, 
          |E^+\rangle \pebra \varphi ^+\rangle \, , 
           \label{ibasivecph+}  \\
       &&\psi ^-=\int_{0}^{\infty} \rmd E \, 
          |E^-\rangle \langle ^-E|\psi ^-\rangle \, . 
           \label{ibasivecph-} 
\end{eqnarray}
The ``in'' (``out'') states can be also expanded in terms of the 
``out'' (``in'') Lippmann-Schwinger bras and kets:
\begin{eqnarray}
       &&\varphi ^+=\int_{0}^{\infty} \rmd E \, 
          |E^-\rangle \langle ^-E| \varphi ^+\rangle \, ,  \\
       &&\psi ^-=\int_{0}^{\infty} \rmd E \, 
          |E^+\rangle \langle ^+E|\psi ^-\rangle \, . 
\end{eqnarray}
By combining Eqs.~(\ref{ibasivecph+})-(\ref{ibasivecph-}) with the relation
\begin{equation}
       \langle ^-E'|E^+\rangle = S(E)\, \delta (E-E') \, ,
\end{equation}
we can express the matrix element~(\ref{iprobaout+-}) as
\begin{equation}
      (\psi ^-,\varphi ^+)=\int_0^{\infty}\rmd E \, 
       \langle \psi ^-|E^-\rangle S(E) \pebra \varphi ^+ \rangle \, .
      \label{psi-ph+interofS}
\end{equation}

Formally as well, the initial and final states can be expanded by the bras 
and kets of the free Hamiltonian:
\begin{eqnarray}
       &&\varphi ^{\rm in} = \int_{0}^{\infty} \rmd E \, 
          |E\rangle \langle E| \varphi ^{\rm in}  \rangle \, , 
            \label{ibasivecphfreein}   \\
       &&\psi ^{\rm out} = \int_{0}^{\infty} \rmd E \, 
          |E\rangle \langle E|\psi ^{\rm out}\rangle \, ,
             \label{ibasivecphfreeout} 
\end{eqnarray}
and the probability amplitude~(\ref{iprobaoutin}) can be written as
\begin{equation}
      (\psi ^{\rm out}, S\varphi ^{\rm in})=\int_0^{\infty}\rmd E \, 
       \langle \psi ^{\rm out}|E\rangle S(E) 
       \langle E|\varphi ^{\rm in} \rangle \, ,
       \label{starpbaisco0II}
\end{equation}
where we have used Eqs.~(\ref{ibasivecphfreein})-(\ref{ibasivecphfreeout}) 
and the relation
\begin{equation}
          \langle E'|S|E\rangle = S(E)\, \delta (E-E') \, .
        \label{smatriel}
\end{equation}
As we shall see in Sec.~\ref{sec:otherreusst}, the formal
expressions~(\ref{ibasivecph+})-(\ref{smatriel}) acquire meaning within the 
RHS.

\section{Solving the Lippmann-Schwinger equation}
\label{sec:LSequation}

For the spherical shell potential, the Lippmann-Schwinger equation can be 
solved explicitly. For the sake of focus, we shall restrict ourselves 
to angular momentum $l=0$.

\subsection{The radial Lippmann-Schwinger equation}

The expression for the spherical shell potential is given by
\begin{equation}
        V({\bf x})\equiv V(r)=\left\{ \begin{array}{ll}
                                0   &0<r<a  \\
                                V_0 &a<r<b  \\
                                0   &b<r<\infty \, ,
                  \end{array} 
                 \right. 
	\label{sbpotential}
\end{equation}
where $V_0$ is a positive number that determines the strength of the 
potential, and $a$ and $b$
determine the positions in between which the potential is non-zero. Because 
the potential~(\ref{sbpotential}) is spherically symmetric, we shall work 
in the radial position representation. In this representation and for $l=0$, 
the free Hamiltonian $H_0$ acts as the formal differential operator $h_0$,
\begin{equation}
       H_0f(r)=h_0f(r)=-\frac{\hbar ^2}{2m}\frac{\rmd ^2}{\rmd r^2}f(r) \, ,
        \label{0doh}
\end{equation}
$V$ acts as multiplication by the rectangular barrier potential $V(r)$,
\begin{equation}
       (Vf)(r)=V(r)f(r) \, ,
\end{equation}
and the total Hamiltonian $H$ acts as the formal differential operator $h$,
\begin{equation}
      Hf(r)=hf(r)=
      \left( -\frac{\hbar ^2}{2m}\frac{\rmd ^2}{\rmd r^2}+V(r) \right) f(r) 
           \, .  \label{doh}
\end{equation}
In the radial representation, the Lippmann-Schwinger equation~(\ref{LSeq1}) 
becomes
\begin{equation}
       \langle r|E ^{\pm}\rangle =\langle r|E\rangle +
       \langle r|\frac{1}{E-H_0\pm \rmi \varepsilon}V|E^{\pm}\rangle \, .
       \label{raLSeq}
\end{equation}
In Eq.~(\ref{raLSeq}), the $\langle r|E\rangle$ are eigenfunctions 
of the formal differential operator $h_0$,
\begin{equation}
      h_0\langle r|E\rangle = E\langle r|E\rangle \, ,
\end{equation}
whereas the $\langle r|E^{\pm}\rangle$ are eigenfunctions of the 
formal differential operator $h$,
\begin{equation}
      h\langle r|E^{\pm}\rangle =E\langle r|E^{\pm}\rangle \, , 
\end{equation}
subject to the boundary conditions specified by 
Eqs.~(\ref{bcoLS1})-(\ref{bcoLS6}).

\subsection{The solutions to the radial Lippmann-Schwinger equation}

The procedure to solve Eq.~(\ref{raLSeq}) is well 
known~\cite{NEWTON,TAYLOR}. Since Eq.~(\ref{raLSeq}) is an integral equation,
it is equivalent to a differential equation subject to the boundary
conditions that are built into that integral equation. In our case,
for $l=0$, Eq.~(\ref{raLSeq}) is equivalent to the Schr\"odinger 
differential equation,
\begin{equation}
      \left( -\frac{\hbar ^2}{2m} \frac{\rmd ^2}{\rmd r^2}+V(r)\right) 
      \langle r|E^{\pm}\rangle =E\langle r|E^{\pm}\rangle \, , 
      \label{schroeque}
\end{equation}
subject to the following boundary conditions:
\numparts
\begin{eqnarray}
      &&\langle r|E^{\pm}\rangle =0 \, ,  \label{bcoLS1} \\
      &&\langle r|E^{\pm}\rangle {\rm \ is \ continuous \ at \ } r=a, b \, , \\
      &&\frac{\rmd}{\rmd r}\langle r|E^{\pm}\rangle 
             {\rm \ is \ continuous \ at \ } r=a, b 
         \, , \\
      && \langle r|E^+\rangle \sim 
          \rme^{-\rmi kr}- S(E) \, \rme ^{\rmi kr} 
                     \quad {\rm as \ } r\to \infty \, ,
        \label{bcoLS5}\\
      && \langle r|E^-\rangle \sim
         \rme ^{\rmi kr} - \overline{S(E)} \,  \rme ^{-\rmi kr} 
             \quad {\rm as \ } r\to \infty \, ,
        \label{bcoLS6}
\end{eqnarray}
\endnumparts
where 
\begin{equation}
      k=\sqrt{\frac{2m}{\hbar ^2}E \, }
      \label{wavenumber}
\end{equation}
is the wave number, and $S(E)$ is the $S$~matrix in the energy 
representation. We recall that the boundary conditions~(\ref{bcoLS5})
and~(\ref{bcoLS6}) originate from the $+\rmi \varepsilon$ and from the 
$-\rmi \varepsilon$
conditions of Eq.~(\ref{raLSeq}), respectively. The boundary 
condition~(\ref{bcoLS5}) means that far away from the potential region,
$\langle r|E^+\rangle$ is a combination of an incoming spherical wave and
an outgoing spherical wave multiplied by the $S$ matrix. The boundary 
condition~(\ref{bcoLS6}) means that far away from the potential region,
$\langle r|E^-\rangle$ is a combination of an outgoing spherical wave and
an incoming spherical wave multiplied by the complex conjugate of the $S$ 
matrix. The asymptotic behaviors~(\ref{bcoLS5}) and~(\ref{bcoLS6}) are the
$l=0$, radial counterparts of the asymptotic behaviors~(\ref{asympbepk}) and
(\ref{asympbemk}).

If we insert expression~(\ref{sbpotential}) for the potential into
Eq.~(\ref{schroeque}), and solve Eq.~(\ref{schroeque}) subject to the 
boundary conditions~(\ref{bcoLS1})-(\ref{bcoLS6}), we obtain the ``in'' 
and ``out'' eigenfunctions,
\begin{equation}
      \langle r|E^{\pm}\rangle \equiv 
       \chi ^{\pm}(r;E)= N(E) \,  
     \frac{\chi (r;E)}{{\cal J}_{\pm}(E)}\, , \qquad E\in [0,\infty ) \, ,
      \label{pmeigndu}
\end{equation}
where $N(E)$ is a delta-normalization factor,
\begin{equation}
      N(E)=\sqrt{\frac{1}{\pi}
         \frac{2m/\hbar ^2}{\sqrt{2m/\hbar ^2 \, E \,}\,}\,} \, ,
        \label{Nfactor}
\end{equation}
$\chi (r;E)$ is the so-called regular solution of Eq.~(\ref{schroeque}),
\begin{equation}
     \hskip-1cm  \chi (r;E) = \left\{
             \begin{array}{ll}
                  \sin ( \sqrt{ \frac{2m}{\hbar^2}E\,}\,r)  & 0<r<a \\
               {\cal J}_1(E)\rme ^{\rmi \sqrt{\frac{2m}{\hbar^2}(E-V_0)\,}\,r}
              +{\cal J}_2(E)\rme^{-\rmi\sqrt{\frac{2m}{\hbar^2}(E-V_0)\,}\,r}
                                                         & a<r<b \\ 
                  {\cal J}_3(E) \rme^{\rmi \sqrt{\frac{2m}{\hbar^2}E\,}\,r}
                   +{\cal J}_4(E) \rme^{-\rmi\sqrt{\frac{2m}{\hbar^2}E\,}\,r}
                                                         & b<r<\infty \, ,
               \end{array}
              \right.
        \label{LSchi}
\end{equation}
and ${\cal J}_{\pm}(E)$ are the Jost functions,
\numparts
\begin{equation}
        {\cal J}_+(E)=-2\rmi {\cal J}_4(E) \, ,
\end{equation}
\begin{equation}
        {\cal J}_-(E)=2\rmi {\cal J}_3(E) \, .
\end{equation}
\endnumparts
The explicit expressions for ${\cal J}_1$-${\cal J}_4$ can be obtained by 
matching the values of $\chi (r;E)$ and of its derivative at the 
discontinuities of the potential
(see Eqs.~(\ref{Jfunctions1})-(\ref{Jfunctions4}) in \ref{sec:A2}). In terms 
of the Jost functions, the $S$ matrix is given by~\cite{NEWTON,TAYLOR}
\begin{equation}
    S(E)=\frac{{\cal J}_-(E)}{{\cal J}_+(E)} \, , \qquad E\in [0,\infty ) \, .
\end{equation}
From Eq.~(\ref{pmeigndu}) it follows that the ``in'' and ``out'' 
Lippmann-Schwinger eigenfunctions are proportional to each other,
\begin{equation}
      \chi ^+(r;E)=S(E) \, \chi ^-(r;E) \, .
         \label{chi+tSe-}
\end{equation}

\subsection{The ``left'' Lippmann-Schwinger eigenfunctions}

When we write Eq.~(\ref{LSeqbraspm}) in the radial position representation, 
we obtain an integral equation for the ``left'' Lippmann-Schwinger
eigenfunctions,
\begin{equation}
       \langle ^{\pm}E| r \rangle =\langle E|r \rangle +
       \langle ^{\pm}E| V\frac{1}{E-H_0\mp \rmi \varepsilon}|r \rangle \, .
       \label{raLSeqbra}
\end{equation}
Solving Eq.~(\ref{raLSeqbra}) is analogous to solving Eq.~(\ref{raLSeq}). For 
$l=0$, Eq.~(\ref{raLSeqbra}) is equivalent to the Schr\"odinger differential 
equation,
\begin{equation}
      \left( -\frac{\hbar ^2}{2m} \frac{\rmd ^2}{\rmd r^2}+V(r)\right) 
      \langle ^{\pm}E|r\rangle =E\langle ^{\pm}E|r\rangle  \, ,
      \label{schroequebra}
\end{equation}
subject to the following boundary conditions:
\numparts
\begin{eqnarray}
      &&\langle ^{\pm}E|0\rangle =0 \, , \label{bcoLS1bra} \\
      &&\langle ^{\pm}E|r\rangle {\rm \ is \ continuous \ at \ } r=a, b \, , \\
      &&\frac{\rmd}{\rmd r}\langle ^{\pm}E|r\rangle {\rm \ is \ 
           continuous \ at \ } r=a, b \, , \\
      && \langle ^+E|r\rangle \sim 
          \rme^{\rmi kr}- \overline{S(E)} \, \rme^{-\rmi kr} 
                \quad {\rm as \ } r\to \infty \, ,
        \label{bcoLS5bra}\\
      && \langle ^-E|r\rangle \sim
         \rme^{-\rmi kr} - S(E) \, \rme^{\rmi kr} 
                 \quad {\rm as \ } r\to \infty \, .
        \label{bcoLS6bra}
\end{eqnarray}
\endnumparts
Note that the boundary conditions~(\ref{bcoLS5bra}) and (\ref{bcoLS6bra})
originate from the $-\rmi \varepsilon$ and from the $+\rmi \varepsilon$ 
conditions of Eq.~(\ref{raLSeqbra}), respectively.

By inserting Eq.~(\ref{sbpotential}) into Eq.~(\ref{schroequebra}), and 
by solving Eq.~(\ref{schroequebra}) subject to the boundary
conditions~(\ref{bcoLS1bra})-(\ref{bcoLS6bra}), we obtain that the ``left''
Lippmann-Schwinger eigenfunctions $\langle ^{\pm}E|r\rangle$ are the complex 
conjugates of the ``right'' Lippmann-Schwinger eigenfunctions 
$\langle r|E^{\pm}\rangle$, 
\begin{equation}
      \langle ^{\pm}E|r\rangle \equiv 
       \overline{\chi ^{\pm}(r;E)} \, .
      \label{pmeigndubra}
\end{equation}

\section{Domain, spectrum and resolvent}
\label{sec:dospresolvent}

In this section, we include the expressions for the domain, the spectrum and 
the resolvent of $H$. Such expressions were obtained in Ref.~\cite{DIS}, and 
are included in this section because we shall need them in subsequent
sections.

The formal differential operator~(\ref{doh}) has an infinite number of 
self-adjoint extensions~\cite{DUNFORD}. These self-adjoint extensions are 
characterized by the following boundary conditions~\cite{DUNFORD}:
\begin{equation}
       f(0)+\alpha  \, f'(0)=0 \, ,
       \quad -\infty < \alpha \leq \infty \, .
        \label{saextensions}
\end{equation}
Among all these, the boundary condition needed in scattering theory is
\begin{equation}
       f(0)=0  \, .
      \label{sac}
\end{equation}
The boundary condition (\ref{sac}) selects the following domain for the
Hamiltonian:
\begin{equation}
     \hskip-2cm  {\cal D}(H) =\left\{ f(r)\, | \ 
             f(r), hf(r)\in L^2([0,\infty ),\rmd r), \,
             f(r) \in AC^2[0,\infty ), \, f(0)=0 \right\}  ,
     \label{LSdomain}
\end{equation}
where $AC^2[0,\infty)$ denotes the space of functions whose first derivative 
is absolutely continuous. This domain induces a self-adjoint operator $H$,
\begin{equation}
      (Hf)(r) = 
      \left( -\frac{\hbar ^2}{2m}\frac{\rmd ^2}{\rmd r^2} +V(r) \right)f(r)
       \, , \quad f \in {\cal D}(H) \, .
      \label{LSoperator}
\end{equation}

The spectrum of $H$, ${\rm Sp}(H)$, was shown in Ref.~\cite{DIS} to 
be $[0,\infty )$. In Ref.~\cite{DIS}, we also calculated the Green 
function $G(r,s;E)$ of $H$ for different regions of the complex energy 
plane. In all our calculations, we used the following branch for the 
square root function:
\begin{equation}
     \hskip-2cm  \sqrt{\, \cdot \ }:
     \left\{ E\in {\mathbb C} \, | \  -\pi <{\rm arg}(E)\leq \pi \right\} 
   \longmapsto \left\{ E\in {\mathbb C} \, | \  
      -\pi/2 <{\rm arg}(E)\leq \pi/2 \right\} 
     \, .
    \label{branchsrfu} 
\end{equation}
The branch~(\ref{branchsrfu}) grants the following relation:
\begin{equation}
      \overline{\sqrt{\overline{E} \, }}= \sqrt{E} \, , \quad 
        E \in \mathbb C \, ,
       \label{cocozc}
\end{equation}
which in general does not hold for other branches of the square root function.

Either by applying Theorem~1 of \ref{sec:A1}, or by borrowing the 
results from Ref.~\cite{DIS}, one can see that in the first quadrant of 
the energy plane, the Green function can be written as
\begin{equation}
      \hskip-2.2cm  G(r,s;E)=\left\{ \begin{array}{ll}
                - \pi  N(E) \, \chi ^+(r;E) \, f^+ (s;E)
               &r<s \\ [2ex]
             - \pi  N(E) \, \chi ^+(s;E) \, f^+ (r;E)
               &r>s  
                  \end{array} 
                 \right.  \quad {\rm Re}(E)>0 \, , \ {\rm Im}(E)> 0 \, ,
      \label{LS++} 
\end{equation}
whereas in the fourth quadrant its expression reads
\begin{equation}
      \hskip-2.2cm  G(r,s;E)=\left\{ \begin{array}{ll}
            - \pi  N(E) \, \chi ^-(r;E) \, f^- (s;E)   
               &r<s \\ [2ex]
          - \pi  N(E) \, \chi ^-(s;E) \, f^- (r;E)
               &r>s  
                  \end{array} 
                 \right. \quad {\rm Re}(E)>0 \, , \ {\rm Im}(E)< 0 \, . 
        \label{LS+-}
\end{equation}
In Eqs.~(\ref{LS++})-(\ref{LS+-}), $\chi ^{\pm}(r;E)$ are given by 
Eq.~(\ref{pmeigndu}), $N(E)$ is given by Eq.~(\ref{Nfactor}), and the 
eigenfunctions $f^{\pm}(r;E)$ are given by
\begin{equation}
     \hskip-2cm
      f^{\pm}(r;E)=\left\{ \begin{array}{lll}
               {\cal A}^{\pm}_1(E)\rme^{\rmi\sqrt{\frac{2m}{\hbar ^2}E\,}\,r}
             +{\cal A}^{\pm}_2(E)\rme^{-\rmi\sqrt{\frac{2m}{\hbar ^2}E\,}\, r} 
                \quad &0<r<a  \\
          {\cal A}^{\pm}_3(E)\rme^{\rmi\sqrt{\frac{2m}{\hbar ^2}(E-V_0)\,}\,r}
       +{\cal A}^{\pm}_4(E)\rme^{-\rmi\sqrt{\frac{2m}{\hbar ^2}(E-V_0)\,}\, r}
                 \quad  &a<r<b \\
               \rme^{\pm \rmi\sqrt{\frac{2m}{\hbar ^2}E\,}\,r}
                 \quad  &b<r<\infty \, . 
               \end{array} 
                 \right. 
            \label{fpmfun}
\end{equation}
The coefficients ${\cal A}^{\pm}_1$-${\cal A}^{\pm}_4$ can be obtained by
matching the values of $f^{\pm}(r;E)$ and of their derivatives at the 
discontinuities of the potential (see 
Eqs.~(\ref{Afunctions3})-(\ref{Afunctions2}) in \ref{sec:A2}).

\section{The operators $U_{\pm}$ and $U_0$}
\label{sec:DIdecomop}

In this section, we obtain the Fourier-like transforms $U_{\pm}$, the
eigenfunction expansions and the direct integral decompositions generated by 
the ``in'' and ``out'' eigenfunctions. The operators $U_{\pm}$ will be 
obtained by applying the Sturm-Liouville theory (see Ref.~\cite{DUNFORD} and 
\ref{sec:A1}). We shall end this section by recalling the expression for the 
Fourier-like transform $U_0$ associated with the free Hamiltonian.

\subsection{The ``in'' unitary operator $U_+$}
\label{sec:DIasink}

The procedure to obtain $U_+$ consists of applying Theorems~2, 3 and 4 of 
\ref{sec:A1}. In order to be able to apply Theorem~4, we 
choose the following basis for the space of solutions of $h\sigma =E\sigma$ 
that is continuous on $(0,\infty )\times (0,\infty )$ and analytically 
dependent on $E$ in a neighborhood of $(0,\infty )$:  
\numparts
\begin{eqnarray}
     \hskip-1cm &&\sigma _1(r;E)=\chi ^+(r;E)\, ,
      \label{LSsigma1=sci} \\ [2ex]
      \hskip-1cm &&\sigma _2(r;E)=\left\{ \begin{array}{lll}
               \cos (\sqrt{\frac{2m}{\hbar ^2}E\,}\, r) \quad &0<r<a  \\
               {\cal C}_1(E)\rme^{\rmi\sqrt{\frac{2m}{\hbar ^2}(E-V_0)\,}\, r}
             +{\cal C}_2(E)\rme^{-\rmi\sqrt{\frac{2m}{\hbar ^2}(E-V_0)\,}\, r}
                 \quad  &a<r<b \\
               {\cal C}_3(E)\rme^{\rmi\sqrt{\frac{2m}{\hbar ^2}E\,}\, r}
                 +{\cal C}_4(E)\rme^{-\rmi\sqrt{\frac{2m}{\hbar ^2}E\,}\, r}
                 \quad  &b<r<\infty \, . 
               \end{array} 
                 \right. \quad
       \label{LSsigam2cos}  
\end{eqnarray}
\endnumparts
The functions ${\cal C}_1$-${\cal C}_4$ can be obtained by matching the values
of $\sigma _2(r;E)$ and its first derivative at the discontinuities of the 
potential (see Eqs.~(\ref{Cfunctions1})-(\ref{Cfunctions4}) in 
\ref{sec:A2}). Equations~(\ref{fpmfun}) and 
(\ref{LSsigma1=sci})-(\ref{LSsigam2cos}) lead to  
\begin{equation}
      f^+(r;E)=\frac{2 \rmi {\cal J}_4(E){\cal C}_4(E)}{N(E)\, W(E)} 
         \sigma _1(r;E)+
      \frac{{\cal J}_4(E)}{W(E)}\sigma _2(r;E) 
      \label{LSThet+inosigm}
\end{equation}
and to
\begin{equation}
      f^-(r;E)=-\frac{2\rmi {\cal J}_4(E){\cal C}_3(E)}{N(E) W(E)} 
       \sigma _1(r;E)-
      \frac{{\cal J}_3(E)}{W(E)}\sigma _2(r;E) \, ,
      \label{LSThet-inosigm}
\end{equation}
where
\begin{equation}
       W(E)\equiv {\cal J}_4(E){\cal C}_3(E)-{\cal J}_3(E){\cal C}_4(E) \, .
      \label{Wdef}
\end{equation}
After substituting Eq.~(\ref{LSThet+inosigm}) into Eq.~(\ref{LS++}) we 
arrive at
\begin{equation}
     \hskip-1cm
   \begin{array}{rl}  
       G(r,s;E)=
      - \pi \, \frac{N(E)}{W(E)} \,  
      \left[ \frac{2\rmi {\cal J}_4(E) {\cal C}_4(E)}{N(E)}
       \sigma _1(r;E)+ {\cal J}_4(E) \sigma _2(r;E)\right] 
       \sigma _1 (s;E) \, ,&  \, \\ [2ex]
     \mbox{Re}(E)>0 \, , \, \mbox{Im}(E)>0\, , \ r>s \, . & \,
   \end{array}     
      \label{LSG++tofpuon}
\end{equation}
After substituting Eq.~(\ref{LSThet-inosigm}) into Eq.~(\ref{LS+-}) we 
arrive at
\begin{equation}
       \hskip-1.6cm
    \begin{array}{rl} 
      G(r,s;E)=
       - \pi  \, \frac{N(E)}{W(E)}
       \frac{{\cal J}_4(E)}{{\cal J}_3(E)}\,
       \left[ \frac{2\rmi {\cal J}_4(E){\cal C}_3(E)}{N(E)}
       \sigma _1(r;E)+{\cal J}_3(E) \sigma _2(r;E)\right]      
       \sigma _1 (s;E) \, , & \, \\ [2ex] 
      \mbox{Re}(E)>0 \, , \, \mbox{Im}(E)<0\, , \ r>s \, , & \,
    \end{array}  
    \label{LSG+-tofpuon}
\end{equation} 
where we have used the relation
\begin{equation}
      \chi ^-(r;E)=-\frac{{\cal J}_4(E)}{{\cal J}_3(E)} \chi ^+(r;E)\, .
          \label{chi-45chip}
\end{equation}
Because by Eq.~(\ref{cocozc})
\begin{equation}
      \overline{\chi ^+(s;\overline{E})} =
      -\frac{{\cal J}_4(E)}{{\cal J}_3(E)}
       \chi ^+(s;E) \, ,
\end{equation}
Eq.~(\ref{LSG++tofpuon}) leads to
\begin{equation}
    \hskip-1.7cm  
    \begin{array}{rl}
       G(r,s;E)=2\pi \rmi \frac{{\cal J}_3(E){\cal C}_4(E)}{W(E)}
       \, \sigma _1(r;E)\overline{\sigma _1(s;\overline{E})}   
      +\pi \frac{N(E){\cal J}_3(E)}{W(E)} \, 
      \sigma _2(r;E)\overline{\sigma _1(s;\overline{E})} \, , &  \\ [2ex]     
        \mbox{Re}(E)>0 \, , \, \mbox{Im}(E)>0 \, , \ r>s \, , & \, 
      \end{array} 
       \label{LSredaot++}
\end{equation}
and Eq.~(\ref{LSG+-tofpuon}) leads to 
\begin{equation}
     \hskip-1.7cm
    \begin{array}{rl}
      G(r,s;E)= 2\pi \rmi \frac{{\cal J}_4(E){\cal C}_3(E)}{W(E)}\,
       \sigma _1(r;E)\overline{\sigma _1(s;\overline{E})}+
      \pi \frac{N(E){\cal J}_3(E)}{W(E)} \, 
       \sigma _2(r;E)\overline{\sigma _1(s;\overline{E})} \, ,  &   \\ [2ex]
       \mbox{Re}(E)>0 \, , \, \mbox{Im}(E)<0 \, , \ r>s \, .& \,
     \end{array} 
       \label{LSredaot+-}
\end{equation}
On the other hand, the Green function can be written in terms of the basis 
$\left\{ \sigma _1,\sigma _2\right\}$ of 
Eqs.~(\ref{LSsigma1=sci})-(\ref{LSsigam2cos}) as (see Eq.~(\ref{Gitofsigma}) 
in Theorem~4)
\begin{equation}
       G(r,s;E)=\sum_{i,j=1}^{2}
       \theta _{ij}^+ (E)\sigma _i(r;E)\overline{\sigma _j(s;\overline{E})}
       \, ,  \qquad r>s \, .
        \label{LSGF++}
\end{equation}
By comparing (\ref{LSGF++}) to (\ref{LSredaot++}), we obtain
\begin{equation}
      \theta _{ij}^+(E)= \left(  \begin{array}{cc}
      2\pi \rmi
      \frac{{\cal J}_3(E){\cal C}_4(E)}{W(E)} 
       & 0 \\
       \pi \frac{N(E){\cal J}_3(E)}{W(E)}
       & 0
                                \end{array}
        \right) , \quad  \mbox{Re}(E)>0 \, , \  \mbox{Im}(E)>0 \, .
       \label{LStheta++}
\end{equation}
By comparing (\ref{LSGF++}) to (\ref{LSredaot+-}), we obtain
\begin{equation}
      \theta _{ij}^+(E)= \left(  \begin{array}{cc}
       2\pi \rmi
       \frac{{\cal J}_4(E) {\cal C}_3(E)}{W(E)}
       & 0 \\
       \pi \frac{N(E){\cal J}_3(E)}{W(E)}
       & 0
                                \end{array}
        \right) , \quad  \mbox{Re}(E)>0 \, , \  \mbox{Im}(E)<0 \, .
      \label{LStheta+-}
\end{equation}
From Eqs.~(\ref{LStheta++}) and (\ref{LStheta+-}) it follows that the measures 
$\rho _{12}$, $\rho _{21}$ and $\rho _{22}$ of Eq.~(\ref{measures}) are zero, 
and that $\rho _{11}$ is given by
\begin{eqnarray}
       \rho _{11}((E_1,E_2))
      &=&\frac{1}{2\pi \rmi} \int_{E_1}^{E_2} 2\pi \rmi 
      \frac{{\cal J}_4(E){\cal C}_3(E)-{\cal J}_3(E){\cal C}_4(E)}{W(E)} \rmd E
       \nonumber \\
      &=&\int_{E_1}^{E_2} \rmd E = E_2-E_1 \, ,
      \label{lebmea}
\end{eqnarray}
where we have used (\ref{Wdef}) in the second step. Therefore, the measure 
$\rho _{11}$ is just the Lebesgue measure, which means, in particular, that
the eigenfunctions $\chi ^+(r;E)$ are $\delta$-normalized (see also Sec.~2.9 
of Ref.~\cite{FP02}).

By Theorem~2 of \ref{sec:A1}, there is a unitary operator
$U_+$ that transforms from the position representation into the energy 
representation,
\begin{equation}
      \hskip-1cm
     \begin{array}{rcl}
      U_+:L^2([0,\infty ),\rmd r) &\longmapsto & L^2([0,\infty ),\rmd E)  
                         \\ [1ex]
       f(r)& \longmapsto & 
         \widehat{f}_+(E)=(U_+f )(E)=\int_0^{\infty}\rmd r f(r) 
                             \overline{\chi ^+(r;E)} \, ,
       \end{array}
      \label{LSinteexpre}
\end{equation}
where $\widehat{f}_+(E)$ denotes the energy
representation of the function $f(r)$ when obtained by way of $U_+$. The 
action of $U_+$ on the domain ${\cal D}(H)$ is given by
\begin{equation}
      \hskip-1.5cm {\cal D}(\widehat{H})=U_+{\cal D}(H)=
                             \left\{ \widehat{f}_+(E) \in 
                             L^2( [0,\infty ),\rmd E) \, | \  
                             \int_0^{\infty}\rmd E \,
                              |E\widehat{f}_+(E)|^2
                             <\infty \right\} .
       \label{LSrhospace}
\end{equation}
It can be easily checked that the operator $U_+$ diagonalizes $H$ in the sense
that $\widehat{H}\equiv U_+HU_+^{-1}$ acts as the operator multiplication 
by $E$,
\begin{equation}
    \begin{array}{rcl}
      \widehat{H}:{\cal D}(\widehat{H}) \subset L^2([0,\infty ),\rmd E)  
        &\longmapsto & L^2([0,\infty ),\rmd E)  \\ [1ex]
        \widehat{f}_+ & \longmapsto & \widehat{H} \widehat{f}_+(E)=
               E\widehat{f}_+(E) \, .
     \end{array}
\end{equation}
The inverse of $U_+$ is given by Eq.~(\ref{inverseU}):
\begin{equation}
      \hskip-1.5cm f(r)=U_+^{-1}\widehat{f}_+(r)= 
        \int_0^{\infty}\rmd E\, \widehat{f}_+(E)\chi ^+(r;E) \, ,
       \quad \widehat{f}_+(E)\in L^2([0,\infty ),\rmd E) \, .
     \label{LSinvdiagonaliza}
\end{equation}
The operator $U_+^{-1}$ transforms from the energy representation back into 
the position representation. Expressions~(\ref{LSinteexpre}) and 
(\ref{LSinvdiagonaliza}) provide the eigenfunction expansions of any square 
integrable function in terms of the ``in'' eigensolutions. 

The operator $U_+$, and therefore the ``in'' eigenfunctions $\chi ^+(r;E)$,
entails a direct integral decomposition of the Hilbert space in a 
straightforward manner:
\begin{equation}
     \begin{array}{rcl}
      {\cal H} &\longmapsto & 
      U_+{\cal H} \equiv 
      \widehat{\cal H}=\oplus \int_{{\rm Sp}(H)}{\cal H}(E)\rmd E  \\ [1ex]
       f &\longmapsto & U_+f\equiv  \left\{ \widehat{f}_+(E) \right\}, 
           \, \quad 
        f\in {\cal D}(H) \, , \, \widehat{f}_+(E) \in {\cal H}(E) \, .
    \end{array}   
    \label{LSdirintdec}
\end{equation}
In this equation, the Hilbert spaces ${\cal H}$, $\widehat{\cal H}$ 
and ${\cal H}(E)$ are respectively realized by $L^2([0,\infty ),\rmd r)$,
$L^2( [0,\infty ),\rmd E)$ and $\mathbb{C}$.

\subsection{The ``out'' unitary operator $U_-$}
\label{sec:outDIasink}

The construction of $U_-$ follows the same procedure as the construction
of $U_+$. The functions
\numparts 
\begin{eqnarray}
      \hskip-1.5cm &&\sigma _1(r;E)=\chi ^-(r;E)\, ,
      \label{-LSsigma1=sci} \\ [2ex]
      \hskip-1.5cm &&\sigma _2(r;E)=\left\{ \begin{array}{lll}
               \cos (\sqrt{\frac{2m}{\hbar ^2}E\,}\, r) \quad &0<r<a  \\
               {\cal C}_1(E)\rme^{\rmi\sqrt{\frac{2m}{\hbar ^2}(E-V_0)\,}\, r}
             +{\cal C}_2(E)\rme^{-\rmi\sqrt{\frac{2m}{\hbar ^2}(E-V_0)\,}\, r}
                 \quad  &a<r<b \\
               {\cal C}_3(E)\rme^{\rmi\sqrt{\frac{2m}{\hbar ^2}E\,}\, r}
                 +{\cal C}_4(E)\rme^{-\rmi\sqrt{\frac{2m}{\hbar ^2}E\,}\, r}
                 \quad  &b<r<\infty 
               \end{array} 
                 \right. \quad
       \label{-LSsigam2cos}  
\end{eqnarray}
\endnumparts
form another basis for the space of solutions of $h\sigma =E\sigma$ that is 
continuous on $(0,\infty )\times (0,\infty )$ and analytically dependent on 
$E$ in a neighborhood of $(0,\infty )$. Therefore, we are allowed to apply 
Theorem~4 of \ref{sec:A1}. Equations~(\ref{fpmfun}) and 
(\ref{-LSsigma1=sci})-(\ref{-LSsigam2cos}) lead to  
\begin{equation}
      f^+(r;E)=-\frac{2\rmi {\cal J}_3(E){\cal C}_4(E)}{N(E)W(E)} 
       \sigma _1(r;E)+\frac{{\cal J}_4(E)}{W(E)}\sigma _2(r;E) 
      \label{-LSThet+inosigm}
\end{equation}
and to
\begin{equation}
      f^-(r;E)=\frac{2\rmi {\cal J}_3(E){\cal C}_3(E)}{N(E)W(E)} 
      \sigma _1(r;E)- \frac{{\cal J}_3(E)}{W(E)}\sigma _2(r;E) \, .
      \label{-LSThet-inosigm}
\end{equation}
By substituting Eqs.~(\ref{-LSThet+inosigm}) and (\ref{chi-45chip}) into 
Eq.~(\ref{LS++}), we obtain
\begin{equation}
    \hskip-2cm  
   \begin{array}{rl}
      G(r,s;E)= \pi \, N(E) \,
      \frac{{\cal J}_3(E)}{{\cal J}_4(E)}  
      \left[ -\frac{2\rmi {\cal J}_3(E) {\cal C}_4(E)}{N(E)W(E)}
       \sigma _1(r;E)+\frac{{\cal J}_4(E)}{W(E)}\sigma _2(r;E)\right] 
       \sigma _1 (s;E) \, , & \, \\ [1ex]
       \mbox{Re}(E)>0 \, , \, \mbox{Im}(E)>0 \, , \ r>s \, . & \,
     \end{array}
        \label{-LSG++tofpuon}
\end{equation}
By substituting Eq.~(\ref{-LSThet-inosigm}) into Eq.~(\ref{LS+-}), we obtain
\begin{equation}
     \hskip-1.5cm  
   \begin{array}{rl}
      G(r,s;E)=-\pi \, N(E) \,
       \left[ \frac{2\rmi {\cal J}_3(E){\cal C}_3(E)}{N(E)W(E)}
       \sigma _1(r;E)-\frac{{\cal J}_3(E)}{W(E)}\sigma _2(r;E)\right]      
       \sigma _1 (s;E) \, , & \, \\ [1ex]
      \mbox{Re}(E)>0 \, , \, \mbox{Im}(E)<0\, , \ r>s \, . & \,
     \end{array} 
      \label{-LSG+-tofpuon}
\end{equation} 
Because by Eq.~(\ref{cocozc})
\begin{equation}
      \overline{\chi ^-(s;\overline{E})}=
      -\frac{{\cal J}_3(E)}{{\cal J}_4(E)}
       \chi ^-(s;E) \, , 
\end{equation}
Eq.~(\ref{-LSG++tofpuon}) leads to
\begin{equation}
     \hskip-1.5cm  
   \begin{array}{rl}
       G(r,s;E)=2\pi \rmi 
       \frac{{\cal J}_3(E){\cal C}_4(E)}{W(E)}
        \sigma _1(r;E)\overline{\sigma _1(s;\overline{E})}   
       -\pi \frac{N(E){\cal J}_4(E)}{W(E)}\sigma _2(r;E)
       \overline{\sigma _1(s;\overline{E})} \, , & \, \\  [1ex]    
      \mbox{Re}(E)>0 \, , \, \mbox{Im}(E)>0\, , \ r>s \, , & \,
     \end{array}
       \label{-LSredaot++}
\end{equation}
and Eq.~(\ref{-LSG+-tofpuon}) leads to 
\begin{equation}
     \hskip-1.5cm  
   \begin{array}{rl}
      G(r,s;E)=2\pi \rmi 
       \frac{{\cal J}_4(E){\cal C}_3(E)}{W(E)}
       \sigma _1(r;E)\overline{\sigma _1(s;\overline{E})}-
        \pi \frac{N(E){\cal J}_4(E)}{W(E)} \sigma _2(r;E)
       \overline{\sigma _1(s;\overline{E})} \, ,  & \, \\ [1ex]
       \mbox{Re}(E)>0 \, , \,  \mbox{Im}(E)<0\, , \ r>s \, .
     \end{array} 
       \label{-LSredaot+-}
\end{equation}
On the other hand, by way of Eq.~(\ref{Gitofsigma}), we can write the Green 
function in terms of the basis 
$\left\{ \sigma _1,\sigma _2 \right\}$ of 
Eqs.~(\ref{-LSsigma1=sci})-(\ref{-LSsigam2cos}) as 
\begin{equation}
       G(r,s;E)=\sum_{i,j=1}^{2}
       \theta _{ij}^+ (E)\sigma _i(r;E)\overline{\sigma _j(s;\overline{E})}
       \, ,   \qquad r>s \, .
        \label{-LSGF++}
\end{equation}
By comparing (\ref{-LSGF++}) to (\ref{-LSredaot++}), we obtain
\begin{equation}
      \theta _{ij}^+(E)= \left(  \begin{array}{cc}
       2\pi \rmi \frac{{\cal J}_3(E){\cal C}_4(E)}{W(E)} 
       & 0 \\
      -\pi \frac{N(E){\cal J}_4(E)}{W(E)}
       & 0
                                \end{array}
        \right) , \quad  \mbox{Re}(E)>0 \, , \  \mbox{Im}(E)>0 \, .
       \label{-LStheta++}
\end{equation}
By comparing (\ref{-LSGF++}) to (\ref{-LSredaot+-}), we obtain
\begin{equation}
      \theta _{ij}^+(E)= \left(  \begin{array}{cc}
      2\pi \rmi \frac{{\cal J}_4(E){\cal C}_3(E)}{W(E)}
       & 0 \\
       -\pi \frac{N(E){\cal J}_4(E)}{W(E)}
       & 0
                                \end{array}
        \right) , \quad  \mbox{Re}(E)>0 \, , \  \mbox{Im}(E)<0 \, .
      \label{-LStheta+-}
\end{equation}
From Eqs.~(\ref{-LStheta++}), (\ref{-LStheta+-}) and (\ref{measures}), it 
follows that the measures $\rho _{12}$, $\rho _{21}$ and $\rho _{22}$ in 
Theorem~4 of \ref{sec:A1} are zero, and that $\rho _{11}$ 
is given by the Lebesgue measure,
\begin{equation}
       \rho _{11}((E_1,E_2))= \int_{E_1}^{E_2} \rmd E =E_2-E_1 \, .
\end{equation}
This means, in particular, that the $\chi ^-(r;E)$ are $\delta$-normalized.

By Theorem~2 of \ref{sec:A1}, there is a unitary operator $U_-$
that transforms from the position into the energy representation,
\begin{equation}
     \hskip-0.5cm \widehat{f}_-(E)=(U_-f)(E)=
       \int_0^{\infty}\rmd r f(r) \overline{\chi ^-(r;E)} \, , \quad
      f(r)\in L^2([0,\infty ),\rmd r) \, ,
      \label{-LSinteexpre}
\end{equation}
where $\widehat{f}_-(E)$ denotes the energy representation of $f(r)$ when
obtained by way of $U_-$. The inverse of $U_-$ is given by 
Eq.~(\ref{inverseU}):
\begin{equation}
     \hskip-1.3cm  f(r)=(U_-^{-1}\widehat{f}_-)(r)= 
        \int_0^{\infty}\rmd E\, \widehat{f}_-(E)\chi ^-(r;E) \, ,
       \quad \widehat{f}_-(E)\in L^2([0,\infty ),\rmd E) \, .
     \label{-LSinvdiagonaliza}
\end{equation}
Likewise $U_+^{-1}$, the operator $U_-^{-1}$ transforms from the energy 
representation into the position representation. Likewise $U_+$, $U_-$ carries
the domain ${\cal D}(H)$ onto the domain~(\ref{LSrhospace}), and 
$\widehat{H}\equiv U_-HU_-^{-1}$ acts as the operator multiplication by 
$E$. As well, expressions~(\ref{-LSinteexpre}) and~(\ref{-LSinvdiagonaliza}) 
provide the expansions of any square integrable function in terms of the
``out'' eigenfunctions, and a direct integral decomposition similar 
to~(\ref{LSdirintdec}).

\subsection{The ``free'' unitary operator $U_0$}
\label{sec:outDIfreeH}

The operator $U_0$ was constructed in Refs.~\cite{DIS,IJTP03}. Because
we shall need $U_0$ in order to construct the M{\o}ller operators, in this 
subsection we recall the expression for $U_0$.

The regular (i.e., vanishing at $r=0$), $\delta$-normalized eigensolution of 
the differential operator~(\ref{0doh}) is given by
\begin{equation}
      \chi _0(r;E)=N(E)\sin \left( \sqrt{\frac{2m}{\hbar ^2}E\,}\,r \right) 
           , \quad 0<r<\infty \, .
      \label{0chi}
\end{equation}
This eigensolution can be used to construct the unitary operator
\begin{equation}
    \begin{array}{rcl}
    U_0:L^2([0,\infty),\rmd r) &\longmapsto & L^2([0,\infty),\rmd E) \\ [1ex]
                              f &\longmapsto & U_0f\equiv {\widehat f}_0
     \end{array}  
\end{equation}
that transforms from the position into the energy representation. The action of 
$U_0$ can be written as an integral operator: 
\begin{equation}
      \widehat{f}_0(E)=U_0f(E)=
       \int_0^{\infty}\rmd r f(r) \overline{\chi _0(r;E)} \, , \quad
      f(r)\in L^2([0,\infty ),\rmd r) \, ,
      \label{0inteexpre}
\end{equation}
where $\widehat{f}_0(E)$ denotes the energy representation of $f(r)$ when
obtained by way of $U_0$. The inverse of $U_0$ can also be written as an 
integral operator:
\begin{equation}
     \hskip-1cm f(r)=U_0^{-1}\widehat{f}_0(r)= 
        \int_0^{\infty}\rmd E\, \widehat{f}_0(E) \chi _0(r;E)   \, ,
       \quad \widehat{f}_0(E)\in L^2([0,\infty ),\rmd E) \, .
     \label{0invdiagonaliza}
\end{equation}
Expressions~(\ref{0inteexpre}) and (\ref{0invdiagonaliza}) provide the
expansions of any square integrable function in terms of the ``free'' 
eigenfunctions.

Note that, since the Lippmann-Schwinger eigenfunctions tend 
to the ``free'' eigenfunctions when the potential vanishes,
\numparts
\begin{equation}
       \lim_{V_0 \to 0} \chi ^{\pm}(r;E)=\chi _0(r;E) \, ,
\end{equation}
the operators $U_{\pm}$ tend to $U_0$ when the potential vanishes, 
\begin{equation}
       \lim_{V_0 \to 0} U_{\pm}=U_0 \, .
\end{equation}
\endnumparts

\section{Construction of the ``in'' and ``out'' bras and kets}
\label{sec:consLSbkin-out}

The solutions to the Lippmann-Schwinger equation are eigenvectors of the 
Hamiltonian whose eigenvalues lie in the continuous spectrum. As explained 
throughly in Refs.~\cite{DIS,EJP05}, eigenvectors 
whose eigenvalues lie in the continuous spectrum must be treated as 
distributions by way of the rigged Hilbert space. In this section, we 
construct the ``in'' and ``out'' bras and kets together with the 
rigged Hilbert spaces that accommodate them. As it turns out, the RHS 
constructed in Refs.~\cite{DIS,JPA02,FP02} suffices for such purpose. The
results of this section will be summarized by Proposition~1 at the end
of this section.

\subsection{The ``in'' kets}

The definition of a ket is borrowed from the theory of 
distributions~\cite{GELFAND}. Given a function $f(x)$ and a space 
of test functions $\mathbf \Phi$, the antilinear functional $F$ associated
with the function $f(x)$ is an integral operator whose kernel is 
precisely $f(x)$:
\begin{equation}
      F(\varphi)=\int \rmd x \, \overline{\varphi (x)} f(x) \, .
    \label{afFtff}
\end{equation}
The ``bad behavior'' of the distribution $f(x)$ must be compensated by the
``nice behavior'' of the test function $\varphi (x)$, so the 
integral~(\ref{afFtff}) makes sense.

By using definition~(\ref{afFtff}), we associate an ``in'' ket $|E^{+} \rangle$
with the ``in'' eigenfunction $\chi ^{+}(r;E)$ for each $E\in [0,\infty )$:
\numparts
\begin{equation}
    \begin{array}{rcl}
       |E^+\rangle :{\mathbf \Phi} & \longmapsto & {\mathbb C}  \\
       \varphi ^+ & \longmapsto & \langle \varphi ^+ |E^+\rangle \equiv
       \int_0^{\infty}\rmd r \, \overline{\varphi ^+(r)} \chi ^+(r;E)  \, .
    \end{array}
        \label{LSdefinitionket+E}
\end{equation}
In Dirac's notation, the action of $|E^+\rangle$ is written as 
\begin{equation}
     \langle \varphi ^+ |E^+\rangle \equiv
       \int_0^{\infty}\rmd r \, \langle \varphi ^+|r\rangle 
      \langle r|E^+\rangle \, .
         \label{LSdefinitionket+EDnotation}
\end{equation}
\endnumparts
Note that even though $\chi ^{+}(r;E) \equiv \langle r|E^+\rangle$ is also 
meaningful for complex energies, the energy in 
Eqs.~(\ref{LSdefinitionket+E})-(\ref{LSdefinitionket+EDnotation}) runs 
only over ${\rm Sp}(H)=[0,\infty )$, because in this paper we restrict 
ourselves to bras and kets associated with energies that belong to the 
spectrum of the Hamiltonian. 

We now need to find the subspace $\mathbf \Phi$ on which 
definition~(\ref{LSdefinitionket+E}) makes sense. Besides making 
(\ref{LSdefinitionket+E}) well defined, the space ${\mathbf \Phi}$ must 
also be invariant under the action of the observables of the system. The 
invariance of $\mathbf \Phi$ is a crucial property, since it entails finite 
expectation values and well-defined commutation relations, and since it
allows us to apply observables on the elements of $\mathbf \Phi$ as many 
times as wished~\cite{EJP05}. Since in this paper the only 
observable we are concerned with is the Hamiltonian, we shall simply require 
invariance under $H$. Thus, the space $\mathbf \Phi$ must satisfy the 
following conditions:
\numparts
 \begin{eqnarray}
  \hskip-1.7cm  &&\bullet \ \, {\rm The \ space \ } {\mathbf \Phi} \ {\rm is 
      \ invariant \
      under \ the \ action \ of } \  H. 
                   \label{condition1} \\ [1ex]
    \hskip-1.7cm  &&\bullet \ \, {\rm The \ elements \ of \  } {\mathbf \Phi} 
    \ {\rm are \ such \ that \ the \ integral \ in \ 
    Eq.~(\ref{LSdefinitionket+E}) \ makes \ sense.}
              \label{condition2} 
 \end{eqnarray}
\endnumparts

In order to meet requirement~(\ref{condition1}), the wave functions 
$\varphi ^+(r)$ must at least be in the maximal invariant subspace of $H$:
\begin{equation}
      {\cal D} = \bigcap_{n=0}^{\infty} {\cal D}(H^n) \, .
\end{equation}
As shown in Refs.~\cite{DIS,FP02}, the space ${\cal D}$ is given by
\begin{equation}
       \begin{array}{ll}
     \hskip-2cm {\cal D}=\left\{ \varphi \in L^2([0,\infty ),\rmd r) \, |\
             \right. &  h^n\varphi (r)\in L^2([0,\infty ),\rmd r),\
       \varphi ^{(n)}(a)=\varphi ^{(n)}(b)=0, \\
       \hskip-2cm \quad & \left. h^n\varphi (0)=0, n=0,1,2,\ldots ; \ 
                \varphi (r) \in C^{\infty}([0,\infty)) \right\}  ,
     \end{array} 
           \label{mainisexi}
\end{equation}
where $\varphi ^{(n)}$ denotes the $n$th derivative of $\varphi$. In order to 
meet requirement~(\ref{condition2}), the wave functions 
$\varphi ^+(r)$ must behave well enough so the integral in 
Eq.~(\ref{LSdefinitionket+E}) is well defined and yields a continuous,
antilinear functional. From the expression for $\chi ^+(r;E)$, 
Eq.~(\ref{pmeigndu}), one can see that the $\varphi ^+(r)$ 
have essentially to control purely imaginary exponentials. Therefore, the 
space $\mathbf \Phi$ constructed in Refs.~\cite{DIS,FP02} meets the 
requirements~(\ref{condition1})-(\ref{condition2}):
\begin{equation}
     \hskip-1.3cm  {\mathbf \Phi}= 
      \left\{ \varphi ^+\in L^2([0,\infty ),\rmd r) \, | \
        \varphi ^+\in {\cal D} \, , 
      \ \| \varphi ^+ \|_{n,m}<\infty \, , \ n,m=0,1,2,\ldots \right\}  ,
         \label{spacePhi}
\end{equation}
where the $\| \ \|_{n,m}$ are given by
\begin{equation}
     \hskip-1.5cm \| \varphi ^+ \|_{n,m} := \sqrt{\int_{0}^{\infty}\rmd r \, 
    \left| (1+r)^n (1+H)^m \varphi ^+(r) 
      \right|^2 \, } \, , \quad n,m=0,1,2, \ldots \, .
      \label{normsLS}
\end{equation}
The space $\mathbf \Phi$ is the collection of square integrable functions 
that belong to the maximal invariant subspace of $H$ and for which the 
estimates~(\ref{normsLS}) are finite. In particular, because $\varphi ^+(r)$ 
satisfies the estimates~(\ref{normsLS}), $\varphi ^+(r)$ falls off at
infinity faster than any polynomial of $r$: 
\begin{equation}
      \lim_{r\to \infty} (1+r)^n\varphi ^+(r) =0 \, , \quad n=0,1,2,\ldots \, .
       \label{zeroinflimit}
\end{equation}
Obviously, the space ${\mathbf \Phi}$ can also be seen
as the maximal invariant subspace of the algebra generated by the Hamiltonian
and the operator multiplication by $r$. The estimates~(\ref{normsLS}) are 
norms (see Proposition~1 at the end of this section), and therefore they 
define a topology (i.e., a meaning of convergence of sequences) 
$\tau _{\mathbf \Phi}$ on $\mathbf \Phi$:  
\begin{equation}
      \varphi ^+_{\alpha}\, 
       \mapupdown{\tau_{\mathbf \Phi}}{\alpha \to \infty}
      \, \varphi ^+\quad {\rm iff} \quad  
      \| \varphi ^+_{\alpha}-\varphi ^+ \| _{n,m} 
      \, \mapupdown{}{\alpha \to \infty}\, 0 \, , \quad n,m=0,1,2, \ldots \, .
\end{equation}   

Once we have constructed the space $\mathbf \Phi$, we can 
construct its topological dual $\mathbf \Phi ^{\times}$ as the 
space of ${\mathbf \Phi}$-continuous antilinear 
functionals on $\mathbf \Phi$, and therewith 
the RHS corresponding to the ``in'' states,
\begin{equation}
      {\mathbf \Phi} \subset L^2([0,\infty ),\rmd r) \subset
      {\mathbf \Phi}^{\times} \, .
      \label{RHSinstates}
\end{equation}
One can show that $|E^+\rangle$ indeed belongs to ${\mathbf \Phi}^{\times}$,
see Proposition~1 below.

The conditions~(\ref{mainisexi}) and (\ref{normsLS}) that determine the
space $\mathbf \Phi$ are very similar to the conditions satisfied by the
Schwartz space on the positive real line, the major difference being that
the derivatives of the elements of $\mathbf \Phi$ vanish at $r=0,a,b$. This 
is why we shall write
\begin{equation}
        {\mathbf \Phi}\equiv \Sw  \, ,
        \label{prospacphe}
\end{equation}
where ${\mathbb R}^+ \equiv [0, \infty )$. With this notation, the 
RHS~(\ref{RHSinstates}) can be written as
\begin{equation}
       \rhsSwt  \, .
      \label{RHSinstates-r-rep}
\end{equation}

The ``in'' kets should be eigenvectors of $H$ as in 
Eq.~(\ref{tisequa}). However, since the Hamiltonian acts in principle only on 
its Hilbert space domain, and since the ``in'' kets belong to the antidual 
rather than to the Hilbert space, we need to extend the 
action of $H$ from $\mathbf \Phi$ to $\mathbf \Phi ^{\times}$, in order
to specify how $H$ acts on $|E^+\rangle$. The theory of distributions 
provides us with a precise prescription for such extension: Given an operator
$A$, the action of $A$ on a functional $|F\rangle \in {\mathbf \Phi}^{\times}$
is defined as
\begin{equation}
     \langle \varphi |A|F\rangle \equiv \langle A^{\dagger}\varphi |F\rangle 
      \, , \quad \mbox{for all} \ \varphi \ \mbox{in} \ \mathbf \Phi  \, .
      \label{adualext}
\end{equation}
It is important to realize that definition~(\ref{adualext}) makes sense only 
when $\mathbf \Phi$ is invariant under $A^{\dagger}$,
\begin{equation}
      A^{\dagger} {\mathbf \Phi} \subset {\mathbf \Phi}  \, .
      \label{invphA}
\end{equation}
Note that when $A$ is self-adjoint (e.g., $A=H$), then $A^{\dagger}=A$, and 
when $A$ is unitary (e.g., $A=U_+$), then 
$A^{\dagger}=A^{-1}$. Definition~(\ref{adualext}) can in turn be used to 
define the notion of eigenket of a self-adjoint observable: A functional 
$|a\rangle$ in $\mathbf \Phi ^{\times}$ is an eigenket of $A$ with 
eigenvalue $a$ if
\begin{equation}
     \langle \varphi |A|a\rangle = \langle A \varphi |a\rangle =
     a \langle \varphi |a\rangle \, , \quad 
    \mbox{for all} \ \varphi \ \mbox{in} \ \mathbf \Phi  \, .
       \label{eeRHS}
\end{equation}
When the ``left sandwiching'' of this equation with the elements of 
$\mathbf \Phi$ is understood and therefore omitted, we shall simply write
\begin{equation}
              A|a\rangle = a |a\rangle \, .
          \label{eeRHSws}
\end{equation}
Thus, within the RHS setting, the eigenvalue equation~(\ref{tisequa}) is to be 
understood as
\begin{equation}
      \langle \varphi ^+ |H|E^+\rangle=
       E\langle \varphi^+|E^+\rangle \, ,
       \quad  \varphi ^+ \in {\mathbf \Phi} \, .
       \label{LSafegenphis}
\end{equation}
This eigenvalue equation is proved in Proposition~1.

By means of the unitary operator $U_+$, which was constructed in 
Section~\ref{sec:DIasink}, we can obtain the energy representation of the space
${\mathbf \Phi}$,
\begin{equation}
      U_+{\mathbf \Phi}= \widehat{\mathbf \Phi}_{+} \, .
\end{equation}
We shall denote the elements of $\widehat{\mathbf \Phi}_{+}$ by 
$\widehat{\varphi}^+(E)\equiv (U_+\varphi ^+)(E)$, where 
$E\in [0, \infty )$. Using the
notation of Eq.~(\ref{prospacphe}), the space $\widehat{\mathbf \Phi}_{+}$ 
will be also denoted as
\begin{equation}
      \widehat{\mathbf \Phi}_{+} \equiv \Swhp \, ,
\end{equation}
and the energy representations of the triplets~(\ref{RHSinstates}) and
(\ref{RHSinstates-r-rep}) will be respectively denoted by
\begin{equation}
       \widehat{\mathbf \Phi}_{+} \subset L^2([0,\infty ),\rmd E)
       \subset  \widehat{\mathbf \Phi}_{+}^{\times} 
\end{equation}
and by
\begin{equation}
      \Swhp \subset  L^2({\mathbb R}^+,\rmd E) \subset \Swhpt \, .
\end{equation}
The antidual extension of $U_+$ yields the energy representation of the 
``in'' kets as
\begin{equation}
      |\widehat{E}^+\rangle \equiv U_+ |E^+\rangle  \, .
\end{equation}
It can be shown that $|\widehat{E}^+\rangle$ acts as the antilinear Schwartz 
delta functional, see Proposition~1.

\subsection{The ``in'' bras}

We now construct the bras $\pebra$ that correspond to the 
kets $|E^+\rangle$. Likewise the definition of a ket, the definition of a bra 
is borrowed from
the theory of distributions~\cite{GELFAND}. Given a function $f(x)$ and
a space of test functions $\mathbf \Phi$, the linear functional $\tilde{F}$
generated by the function $f(x)$ is an integral operator whose kernel
is the complex conjugate of $f(x)$:
\begin{equation}
      \tilde{F}(\varphi)=\int \rmd x \, \varphi (x) \overline{f(x)} \, .
           \label{afGtff}
\end{equation}
By using prescription~(\ref{afGtff}), we define the bra $\pebra$ associated
with the ``in'' eigenfunction $\chi ^+(r;E)$ as
\numparts
\begin{equation}
     \pebra \varphi ^+\rangle \equiv
       \int_0^{\infty} \rmd r \, \overline{\chi^+(r;E)} \varphi ^+(r)  \, ,
      \label{defibraEplus}
\end{equation}
which in Dirac's notation becomes
\begin{equation}
     \pebra \varphi ^+\rangle \equiv
       \int_0^{\infty} \rmd r \, 
              \langle ^+E|r\rangle \langle r|\varphi ^+\rangle  \, .
      \label{braerpilsk}
\end{equation}
\endnumparts
Here $\langle ^+E|r\rangle$ denotes the complex conjugate of $\chi ^+(r;E)$. 

From definitions~(\ref{LSdefinitionket+E}) and (\ref{defibraEplus}), it
follows that the action of the bras $\pebra$ is complex conjugated to the
action of the kets $|E^+ \rangle$:
\begin{equation}
      \pebra \varphi ^+ \rangle =
       \overline{\langle \varphi ^+|E^+ \rangle}= \widehat{\varphi}^+(E)\, .
        \label{baccka}
\end{equation}
In its turn, Eq.~(\ref{baccka}) shows that definition~(\ref{defibraEplus})
is meaningful when $\varphi ^+ \in {\mathbf \Phi}$. If we denote by
${\mathbf \Phi}^{\prime}$ the space of continuous {\it linear} functionals 
over ${\mathbf \Phi}$, then it is also easy to prove that $\pebra$ belongs to 
${\mathbf \Phi}^{\prime}$, see Proposition~1 below. Therefore, the triplet
\begin{equation}
      {\mathbf \Phi} \subset L^2([0,\infty ),\rmd r) \subset 
          {\mathbf \Phi}^{\prime} 
       \label{tripletprme}
\end{equation}
is suitable to accommodate the ``in'' bras. Using the notation
introduced in Eq.~(\ref{prospacphe}), we shall denote the 
triplet~(\ref{tripletprme}) as
\begin{equation}
      \Sw \subset L^2({\mathbb R}^+,\rmd r) \subset 
          \Swp \, . 
       \label{tripletprmer-rep}
\end{equation}

Our next task is showing that the ``in'' bras are left eigenvectors of 
the Hamiltonian. For this purpose, we need to specify how $H$ acts on the 
bras, that is, how $H$ acts on the dual space ${\mathbf \Phi}^{\prime}$. The 
action to the left of an operator $A$ on a linear functional 
$\langle F| \in {\mathbf \Phi}^{\prime}$ is defined as
\begin{equation}
     \langle F|A|\varphi \rangle \equiv \langle F| A^{\dagger}\varphi \rangle 
      \, , \quad \mbox{for all} \ \varphi \ \mbox{in} \ \mathbf \Phi  \, .
      \label{dualext}
\end{equation}
In turn, Eq.~(\ref{dualext}) can be used to define the notion of eigenbra of 
a self-adjoint operator: A functional $\langle a|$ in 
$\mathbf \Phi ^{\prime}$ is an eigenbra of $A$ with eigenvalue $a$ if
\begin{equation}
     \langle a |A|\varphi \rangle = \langle a|A \varphi \rangle =
     a \langle a|\varphi \rangle \, , \quad 
    \mbox{for all} \ \varphi \ \mbox{in} \ \mathbf \Phi  \, .
       \label{eeRHSb}
\end{equation}
When the ``right sandwiching'' of this equation with the elements of 
$\mathbf \Phi$ is understood and therefore omitted, we shall simply write
\begin{equation}
           \langle a |A = a \langle a|  \, .
         \label{eeRHSwsb}
\end{equation}
The eigenbra equation~(\ref{LSeibenbraeq}) is therefore to be understood as
\begin{equation}
      \pebra H|\varphi ^+\rangle 
       =E\pebra \varphi ^+\rangle \, , \quad \varphi ^+ \in \Sw \, .
            \label{phikpssleftkeofH}
\end{equation}
Proposition~1 below proves this equation.

The operator $U_+$ can also be extended to the dual space, and such extension
can be used to obtain the energy representation of $\pebra$,
\begin{equation}
      \langle ^+\widehat{E}|= \langle ^+E|U_+ \, .
\end{equation}
In Proposition~1, we show that $\langle ^+\widehat{E}|$ acts as the linear 
Schwartz delta functional.

It should be emphasized that there is a one-to-one correspondence between 
``in'' bras and ``in'' kets, since to each ``in'' ket $|E^+\rangle$ there 
corresponds an ``in'' bra $\pebra$, and vice versa.

\subsection{The ``out'' kets}

The construction of the ``out'' kets closely parallels the construction of 
the ``in'' kets. By way of~(\ref{afFtff}), we define the ``out'' kets 
$|E^{-} \rangle$ for each $E\in [0,\infty )$ as
\numparts
\begin{equation}
    \begin{array}{rcl}
       |E^-\rangle :{\mathbf \Phi} & \longmapsto & {\mathbb C}  \\
       \psi ^- & \longmapsto & \langle \psi ^- |E^-\rangle \equiv
       \int_0^{\infty}\rmd r \, \overline{\psi ^-(r)} \chi ^-(r;E)  \, ,
    \end{array}
        \label{outLSdefinitionket+E}
\end{equation}
where $\chi ^{-}(r;E)$ is the ``out'' eigenfunction of 
Eq.~(\ref{pmeigndu}). In Dirac's notation, the action of $|E^-\rangle$ reads
as
\begin{equation}
    \langle \psi ^- |E^-\rangle \equiv
       \int_0^{\infty}\rmd r \, \langle \psi ^-|r\rangle 
      \langle r|E^-\rangle \, .
         \label{outLSdefinitionket+EDnotation}
\end{equation}
\endnumparts

Similarly to the ``in'' case, the space $\mathbf \Phi$ in
definition~(\ref{outLSdefinitionket+E}) must satisfy the following conditions:
\numparts
 \begin{eqnarray}
  \hskip-2.2cm  &&\bullet \ \, {\rm The \ space \ } {\mathbf \Phi} \ {\rm is 
      \ invariant \
      under \ the \ action \ of } \  H. 
                   \label{outcondition1} \\ [1ex]
    \hskip-2.2cm  &&\bullet \ \, {\rm The \ elements \ of \  } {\mathbf \Phi} 
    \ {\rm are \ such \ that \ the \ integral \ in \ 
    Eq.~(\ref{outLSdefinitionket+E}) \ makes \ sense.}
              \label{outcondition2} 
 \end{eqnarray}
\endnumparts
Likewise the $\chi ^+(r;E)$, the $\chi ^-(r;E)$ behave like purely imaginary
exponentials when $E\in [0,\infty )$, which implies that
conditions~(\ref{outcondition1})-(\ref{outcondition2}) are equivalent
to conditions~(\ref{condition1})-(\ref{condition2}). Therefore, the space of
``out'' wave functions is the same as the space of ``in'' wave functions:
\begin{equation}
     \hskip-1.3cm  {\mathbf \Phi}= 
      \left\{ \psi ^-\in L^2([0,\infty ),\rmd r) \, | \
        \psi ^-\in {\cal D} \, , 
      \ \| \psi ^- \|_{n,m}<\infty \, , \ n,m=0,1,2,\ldots \right\}  ,
         \label{outspacePhi}
\end{equation}
where the $\| \ \|_{n,m}$ are given by Eq.~(\ref{normsLS}). It can be proved 
that the ``out'' kets $|E^-\rangle$ belong to $\Swt$, and 
that they are eigenvectors of $H$, see Proposition~1 below.

A comment on notation is in order here. We have used two different symbols,
$\varphi ^+$ and $\psi ^-$, to denote the elements of one and the same 
space $\mathbf \Phi$, see Eqs.~(\ref{outspacePhi}) and (\ref{spacePhi}). The 
reason why we use two different symbols is that we need to specify what kets
are acting on the elements of $\mathbf \Phi$. When the elements of 
$\mathbf \Phi$ are acted upon by $|E^+\rangle$ ($|E^-\rangle$), we shall use 
the notation $\langle \varphi ^+|E^+\rangle$ 
($\langle \psi ^-|E^-\rangle$). Another reason why we need two symbols
is that, as we shall explain in Sec.~\ref{sec:timevoLSbk}, the time 
evolution of $\varphi ^+$ is interpreted in a different way to the time
evolution of $\psi ^-$. 

By means of the unitary operator $U_-$, which was constructed in 
Section~\ref{sec:outDIasink}, we can obtain another energy representation of 
the space ${\mathbf \Phi}$,
\begin{equation}
      U_-{\mathbf \Phi}= \widehat{\mathbf \Phi}_{-} \, .
\end{equation}
We shall denote the elements of $\widehat{\mathbf \Phi}_{-}$ as 
$\widehat{\psi}^-(E)\equiv (U_-\psi ^-)(E)$, where 
$E\in [0, \infty )$. Using the
notation of Eq.~(\ref{prospacphe}), the space $\widehat{\mathbf \Phi}_{-}$ 
will be also denoted as
\begin{equation}
      \widehat{\mathbf \Phi}_{-} \equiv \Swhn \, ,
\end{equation}
and the energy representations of the triples~(\ref{RHSinstates}) and
(\ref{RHSinstates-r-rep}), when obtained through $U_-$, will be respectively 
denoted by
\begin{equation}
       \widehat{\mathbf \Phi}_{-} \subset L^2([0,\infty ),\rmd E)
       \subset  \widehat{\mathbf \Phi}_{-}^{\times} 
\end{equation}
and by
\begin{equation}
      \Swhn \subset  L^2({\mathbb R}^+,\rmd E) \subset \Swhnt \, .
\end{equation}
As well, the dual extension of $U_-$ yields the energy representation of the 
``out'' kets,
\begin{equation}
        |\widehat{E}^-\rangle \equiv U_- |E^-\rangle \, .
\end{equation}
In Proposition~1, we prove that $|\widehat{E}^-\rangle$ acts as the
antilinear Schwartz delta functional.

\subsection{The ``out'' bras}

By using prescription~(\ref{afGtff}), we define the bra $\nebra$ as
\numparts
\begin{equation}
     \nebra \psi ^-\rangle \equiv
       \int_0^{\infty} \rmd r \, \overline{\chi^-(r;E)} \psi ^-(r)  \, ,
      \label{outdefibraEplus}
\end{equation}
which in Dirac's notation becomes
\begin{equation}
     \nebra \psi ^-\rangle \equiv
       \int_0^{\infty} \rmd r \, 
              \langle ^-E|r\rangle \langle r|\psi ^-\rangle  \, ,
      \label{outbraerpilsk}
\end{equation}
\endnumparts
where $\langle ^-E|r\rangle$ denotes the complex conjugate of $\chi ^-(r;E)$. 

From definitions~(\ref{outLSdefinitionket+E}) and (\ref{outdefibraEplus}), it
follows that the action of the bras $\nebra$ is complex conjugated to the
action of the kets $|E^- \rangle$:
\begin{equation}
      \nebra \psi ^- \rangle =
       \overline{\langle \psi ^-|E^- \rangle}= \widehat{\psi}^-(E)\, .
        \label{outbaccka}
\end{equation}
In turn, Eq.~(\ref{outbaccka}) shows that definition~(\ref{outdefibraEplus}) 
is meaningful when $\psi ^- \in {\mathbf \Phi}$. Therefore, the ``in''
bras belong to $\Swp$, and they are left eigenvectors of $H$. Also,
the energy representation of $\nebra$, 
$\langle ^-\widehat{E}| \equiv \nebra U_-$, acts as the linear Schwartz delta
functional.

The following proposition summarizes the results of this section:

\vskip0.5cm

\newtheorem*{Prop1}{Proposition~1}
\begin{Prop1} \label{Prop1} The triplets of spaces (\ref{tripletprmer-rep}) and
(\ref{RHSinstates-r-rep}) are rigged Hilbert spaces, and they satisfy all 
the requirements needed to accommodate the Lippmann-Schwinger bras and 
kets. More specifically,
\begin{itemize}
\item[({\it i})] The $\| \ \|_{n,m}$ are norms.

\item[({\it ii})] The space $\Sw$ is dense in
$L^2([0,\infty ),\rmd r)$.

\item[({\it iii})] The space $\Sw$ is invariant under
the action of the Hamiltonian, and $H$ is ${\mathbf \Phi}$-continuous.

\item[({\it iv})] The kets $|E^{\pm}\rangle$ are antilinear 
functionals over $\Sw$, i.e., $|E^{\pm}\rangle \in \Swt$. 

The bras $\langle ^{\pm}E|$ are linear functionals over $\Sw$, 
i.e., $\langle ^{\pm}E| \in \Swp$.

\item[({\it v})] The kets $|E^{\pm}\rangle$ are (right) eigenvectors of $H$ 
with eigenvalue $E$,
\begin{equation}
          H|E^{\pm}\rangle=E|E^{\pm}\rangle \, .
       \label{keigeeq}
\end{equation}
The bras $\langle ^{\pm}E|$ are (left) eigenvectors of $H$ with
eigenvalue $E$,
\begin{equation}
      \langle ^{\pm}E| H=E\langle ^{\pm}E|   \, .
        \label{kpssleftkeofH}
\end{equation}

\item[({\it vi})] In the energy representation, the ``in'' and ``out'' kets 
act as the antilinear Schwartz delta functional:
\begin{equation}
        \langle \widehat{\varphi}^+|\widehat{E}^+\rangle= 
           \overline{\widehat{\varphi}^+(E)} \, ,
       \label{aSdf}
\end{equation}
\begin{equation}
        \langle \widehat{\psi}^-|\widehat{E}^-\rangle= 
           \overline{\widehat{\psi}^-(E)} \, ,
       \label{lSdf}
\end{equation}
whereas the ``in'' and ``out'' bras act as the linear Schwartz delta 
functional:
\begin{equation}
      \langle ^+\widehat{E}|\widehat{\varphi}^+\rangle = \widehat{\varphi}^+(E)
        \, ,
   \label{pluslSdf}
\end{equation}
\begin{equation}
      \langle ^-\widehat{E}|\widehat{\psi}^-\rangle = \widehat{\psi}^-(E)
        \, .
   \label{minuslSdf}
\end{equation}
\end{itemize}
\end{Prop1}

\vskip0.5cm

The proof of this proposition can be found in~\ref{sec:A3}.

\section{The time evolution of the Lippmann-Schwinger bras and kets}
\label{sec:timevoLSbk}

In the previous sections, we obtained the (time-independent) solutions to the 
Lippmann-Schwinger equations. In this section, we obtain the time evolution of 
$|E^{\pm}\rangle$, $\langle ^{\pm}E|$, $\varphi ^+$ and $\psi ^-$.

\subsection{The time evolution of the ``in'' states and kets}

In Quantum Mechanics, time evolution follows from the action of the 
operator $\rme^{-\rmi Ht/\hbar}$. This operator is unitary for each 
$t \in (-\infty ,\infty )$, and the set 
\begin{equation}
      \left\{ \rme^{-\rmi Ht/\hbar} \, | \ -\infty <t<\infty \right\}
      \label{groupeevloII}
\end{equation}
is a one-parameter unitary group. For each instant $t$, the time evolution 
of the ``in'' states is then given by
\begin{equation}
     \varphi ^+(r;t)=\rme ^{-\rmi Ht/\hbar }\varphi ^+(r) \, , \quad  
        t \in {\mathbb R}  \, . 
\end{equation}
This time evolution can be conveniently written in terms of the ``in'' 
eigenfunctions by formally applying $\rme ^{-\rmi Ht/\hbar }$ to both sides
of the expansion~(\ref{LSinvdiagonaliza}):
\begin{equation}
      \left( \rme^{-\rmi Ht/\hbar}\varphi ^+\right) (r)=
       \int_0^{\infty} \rmd E\, \rme^{-\rmi Et/\hbar}
       \widehat{\varphi}^+(E) \chi ^+(r;E)  \, ,
       \label{timevolwispprII} 
\end{equation}
which in Dirac's notation becomes
\begin{equation}
      \langle r |\rme^{-\rmi Ht/\hbar}|\varphi ^+ \rangle =
       \int_0^{\infty} \rmd E\, \rme^{-\rmi Et/\hbar}
       \langle r|E ^+\rangle  \langle ^+E| \varphi ^+ \rangle   \, .
       \label{timevolwispprIIDirac} 
\end{equation}
Equation~(\ref{timevolwispprII}) is equivalent to defining
$\rme^{-\rmi Ht/\hbar}$ as the operator that in the energy representation
acts as multiplication by $\rme^{-\rmi Et/\hbar}$:
\begin{equation}
      \left( U_+\rme^{-\rmi Ht/\hbar}\varphi ^+  \right) (E) =
      \left( \rme^{-\rmi \widehat{H}t/\hbar}\widehat{\varphi}^+\right) (E)=
       \rme^{-\rmi Et/\hbar} \widehat{\varphi}^+(E) \, . 
      \label{timevolwispprerII}
\end{equation}
Expressions~(\ref{timevolwispprII}) and (\ref{timevolwispprerII}) can be
rigorously justified by way of Eq.~(\ref{Gfunction}) of Theorem~2.

In order to obtain the time evolution of the ``in'' kets $|E^+\rangle$, we 
need to extend $\rme^{-\rmi Ht/\hbar}$ to the antidual space. Such extension 
follows from prescription~(\ref{adualext}):
\begin{equation}
       \langle \varphi ^+|\rme ^{-\rmi Ht/\hbar}|E^+\rangle :=
      \langle \rme ^{\rmi Ht/\hbar}\varphi ^+|E^+\rangle \, , \quad
      \varphi ^+ \in \Sw \, .
         \label{timevolinkets}
\end{equation} 
As noted in Sec.~\ref{sec:consLSbkin-out}, this definition makes sense only 
when $\Sw$ is invariant under $\rme ^{\rmi Ht/\hbar}$, 
\begin{equation}
       \rme ^{\rmi Ht/\hbar} \Sw \subset \Sw  \, .
         \label{invaraincS}
\end{equation}
Such invariance is guaranteed for all $t$ by Hunziker's 
theorem~\cite{HUNZIKER}, see the theorem below, and therefore the time 
evolution of the ``in'' kets is well defined for all $t$. We note in passing 
that, as explained in Ref.~\cite{COCOYOC}, the invariance of the space of 
test functions is equivalent to having a well-defined Heisenberg picture, and 
therefore, from a physical point of view, it is clear that such invariance 
must hold.

In order to state Hunziker's 
theorem, we need some definitions: A potential $V$ is said to satisfy the 
Kato condition~\cite{KATO} if ${\cal D}(H_0) \subset {\cal D}(V)$, and if 
there exist constants $a<1$, $b<\infty$, such that, for all 
$f \in {\cal D}(H_0)$,
\begin{equation}
    \| Vf \| \leq a \|H_0 \| + b \|f\| \, .
        \label{katocond}
\end{equation}
When $V$ satisfies~(\ref{katocond}), then $V$ can be seen as a small 
perturbation to the kinetic energy. For any positive integer $n$, we 
define a linear subset $D_n$ of $L^2({\mathbb R}^+,\rmd r)$ and a norm
$\| \ \|_n$ on $D_n$ by
\begin{equation}
      D_n:=\bigcap_{\Sb k\leq n \\ m\leq n-k \endSb} {\cal D}(r^kH^m) \, , 
\hskip2cm
      \|f\|_n:= \sup_{\Sb k\leq n \\ m\leq n-k \endSb}\|r^kH^mf\| \, , 
\end{equation}
where $m$ is a positive integer.

\vskip0.5cm

\newtheorem*{ThH}{Theorem}
\begin{ThH} \label{Th1} (Hunziker) When the potential $V$ satisfies the
Kato condition~(\ref{katocond}), then the following holds for any 
positive integer $n$:
\begin{itemize}
      \item[(i)] $D_n$ is invariant under the unitary group 
                 $\rme ^{-\rmi Ht/\hbar}$.
      \item[(ii)] For any $f\in D_n$, $\rme ^{-\rmi Ht/\hbar}f$ is continuous
in $t$ in the sense of the norm $\| \ \|_n$, and there exists a constant
$c_n$ such that 
   \begin{equation}
       \| \rme ^{-\rmi Ht/\hbar}f \|_n \leq 
             c_n \left( 1+|t|/\hbar \right) ^n \| f \|_n \, .
   \end{equation}
       \item[(iii)] For any $f\in D_n$,
   \begin{equation}
        r^n\rme ^{-\rmi Ht/\hbar}f= \rme ^{-\rmi Ht/\hbar}r^nf+
         \frac{\rmi}{\hbar}
        \int_0^t\rmd \tau \, \rme ^{-\rmi H(t-\tau)/\hbar} \, [H,r^n] \,
           \rme ^{-\rmi H\tau/\hbar}f \, ,
    \end{equation}
the commutator being defined as
\begin{equation}
      [H,r^n]=- \frac{\rmi \hbar}{m} \dot{r}^nP
               -\frac{\hbar ^2}{2m}\ddot{r}^n \, ,
\end{equation}
where $\dot{r}^n$ and $\ddot{r}^n$ denote the first and second derivatives
of $r^n$ with respect to $r$, and where 
$P\equiv -\rmi \hbar\, \rmd /\rmd r$. In the $L^2$-norm, the integrand is 
continuous in $\tau$ and bounded by 
${\rm constant}\, (1+|\tau|/\hbar )^{n-1} \|f\|_n$.
\end{itemize}
\end{ThH}

\vskip0.3cm

We note that Hunziker's theorem, as stated in Ref.~\cite{HUNZIKER},
is only valid for Hamiltonians defined on $L^2({\mathbb R}^N,\rmd x)$ for any 
dimension $N$. We therefore have to adapt the proof of Hunziker's theorem to 
our case, which involves a Hamiltonian defined on
$L^2({\mathbb R}^+,\rmd r)$. Such adaptation will not be reproduced here,
since it is straightforward.

Our rectangular barrier potential clearly satisfies the Kato condition. Hence, 
Hunziker's theorem applies to our case. Because
$\Sw = \bigcap_{n=0}^{\infty} D_n$, and because Hunziker's theorem 
guarantees that each $D_n$ is invariant under $\rme ^{-\rmi Ht/\hbar}$, the 
invariance~(\ref{invaraincS}) holds, and hence definition~(\ref{timevolinkets})
makes sense.

After seeing that it makes sense, we are going to see that 
definition~(\ref{timevolinkets}) yields the expected time evolution of
the ``in'' kets:
\begin{equation}
      \rme ^{-\rmi Ht/\hbar}|E^{+}\rangle = 
           \rme ^{-\rmi Et/\hbar} |E^{+}\rangle \, , 
           \quad -\infty < t < \infty  \, ,
        \label{timevolinketsH}
\end{equation}
which in the RHS language is to be understood as
\begin{equation}
    \langle \varphi ^+| \rme ^{-\rmi Ht/\hbar}|E^{+}\rangle = 
           \rme ^{-\rmi Et/\hbar} \langle \varphi ^+|E^{+}\rangle \, ,
     \quad  \varphi ^+ \in \Sw \, .
    \label{timevoinket}
\end{equation}
Equation~(\ref{timevoinket}) follows from the following chain of equalities:
\begin{eqnarray}
       \langle \varphi ^+| \rme ^{-\rmi Ht/\hbar}|E^{+}\rangle 
       &=\langle \rme ^{\rmi Ht/\hbar} \varphi ^+|E^{+}\rangle 
         \hskip2cm
           &\mbox{by definition~(\ref{timevolinkets})}  \nonumber   \\
        &=  \overline{\rme ^{\rmi \widehat{H}t/\hbar} \widehat{\varphi}^+(E)}
                &\mbox{by Eq.~(\ref{timevolwispprerII})} 
                         \nonumber   \\  
        &= \rme ^{-\rmi Et/\hbar} \overline{\widehat{\varphi}^+(E)} 
             &\mbox{by Eq.~(\ref{timevolwispprerII})}  
             \nonumber   \\
        &= \rme ^{-\rmi Et/\hbar} \langle \varphi ^+|E^{+}\rangle   \, .
            & \quad 
\end{eqnarray}

Note that in the Hardy-function approach to the Lippmann-Schwinger equation,
the time evolution of the Lippmann-Schwinger bras and kets is not defined
for all times, but only for positive (or negative) times. Also, that
the time evolution~(\ref{timevolinketsH}) is valid for all times could have
been anticipated from the physics of a scattering process: The ``in'' solution
of the Lippmann-Schwinger equation represents a monoenergetic ingoing particle 
prepared in the distant past ($t=-\infty$) that hits the target and evolves 
into an outgoing particle in the distant future ($t=+\infty$). Thus, 
physically, the process described by the ``in'' ket lasts from $t=-\infty$
until $t=+\infty$, in agreement with~(\ref{timevolinketsH}) but in 
disagreement with the Hardy-function approach.

\subsection{The time evolution of the ``in'' bras}

The time evolution of $\langle ^+E|$ can be obtained by extending
$\rme ^{-\rmi Ht/\hbar}$ to the dual space. Such
extension follows from definition~(\ref{dualext}):
\begin{equation}
       \langle ^+E| \rme ^{-\rmi Ht/\hbar}|\varphi ^+ \rangle :=
      \langle ^+E| \rme ^{\rmi Ht/\hbar}\varphi ^+ \rangle \, , \quad
      \varphi ^+ \in \Sw \, .
         \label{timevolinbras}
\end{equation} 
Likewise definition~(\ref{timevolinkets}), definition~(\ref{timevolinbras}) 
makes sense because $\Sw$ is invariant under $\rme ^{\rmi Ht/\hbar}$.

By using Eqs.~(\ref{baccka}) and (\ref{timevoinket}), it can be easily 
seen that definition~(\ref{timevolinbras}) yields
\begin{equation}
       \langle ^+E| \rme ^{-\rmi Ht/\hbar}|\varphi ^+ \rangle =
         \rme ^{\rmi Et/\hbar} \langle ^+E|\varphi ^+ \rangle  \, ,
        \quad -\infty < t < \infty  \, ,
         \label{timevolinbras1}
\end{equation} 
which, after omitting the $\varphi ^+$, becomes the expected result:
\begin{equation}
   \langle ^+E|\, \rme ^{-\rmi Ht/\hbar} = \rme ^{\rmi Et/\hbar} \langle ^+E|  
          \, ,
    \quad -\infty < t < \infty  \, .
         \label{timevolinbras2}
\end{equation}

\subsection{The time evolution of the ``out'' states, bras and kets}
\label{sec:tidesolseqoutII}

The time evolution of the ``out'' states can be obtained by formally
applying $\rme^{-\rmi Ht/\hbar}$ to both sides of 
Eq.~(\ref{-LSinvdiagonaliza}):
\begin{equation}
      \left( \rme^{-\rmi Ht/\hbar}\psi ^-\right) (r)=
       \int_0^{\infty} \rmd E\, \rme^{-\rmi Et/\hbar}
       \widehat{\psi}^-(E) \chi ^-(r;E)  \, ,
       \label{timevolwispproutII} 
\end{equation}
which in Dirac's notation becomes
\begin{equation}
      \langle r| \rme^{-\rmi Ht/\hbar}| \psi ^-\rangle =
       \int_0^{\infty} \rmd E\, \rme^{-\rmi Et/\hbar}
       \langle r|E^-\rangle  \langle ^-E| \psi ^- \rangle  \, .
       \label{timevolwispproutIIDirac} 
\end{equation}
Likewise Eq.~(\ref{timevolwispprII}), Eq.~(\ref{timevolwispproutII}) is
tantamount to defining $\rme^{-\rmi Ht/\hbar}$ as the operator that in 
the energy representation acts as multiplication by $\rme^{-\rmi Et/\hbar}$:
\begin{equation}
      \left( U_-\rme^{-\rmi Ht/\hbar}\psi ^-  \right) (E) =
      \left( \rme^{-\rmi \widehat{H}t/\hbar}\widehat{\psi}^-\right) (E)=
       \rme^{-\rmi Et/\hbar} \widehat{\psi}^-(E) \, . 
      \label{timevolwisppreroutII}
\end{equation}
Likewise expressions~(\ref{timevolwispprII}) and (\ref{timevolwispprerII}),
expressions~(\ref{timevolwispproutII}) and (\ref{timevolwisppreroutII}) can be
rigorously justified by way of Eq.~(\ref{Gfunction}) of Theorem~2.

By using Eq.~(\ref{timevolwispproutII}), and following the same steps as
for the ``in'' bras and kets, one can easily obtain the time evolution of 
the ``out'' bras and kets:
\begin{eqnarray}
   \langle ^-E|\, \rme ^{-\rmi Ht/\hbar} = \rme ^{\rmi Et/\hbar} \langle ^-E| 
           \, , \quad -\infty < t < \infty  \, ,   \\ [1ex]
         \rme ^{-\rmi Ht/\hbar}|E^{-}\rangle = 
           \rme ^{-\rmi Et/\hbar} |E^{-}\rangle \, , 
           \quad -\infty < t < \infty  \, .
\end{eqnarray}

As already mentioned in Sec.~\ref{sec:consLSbkin-out}, the time evolution of 
the ``in'' states has a different interpretation from the time evolution of the
``out'' states. By the stationary-phase method, one can easily see that the
``in'' states are determined by the {\it initial} (i.e., prepared) 
condition that, as $t \to -\infty$ and $r\to \infty$, they move toward the 
potential region, whereas the ``out'' states are determined by the 
{\it final} (i.e., detected) condition that, as $t \to \infty$ and 
$r\to \infty$, they move away from the 
potential region (see also Ref.~\cite{RSIII}, p.~356). This is why
we have denoted the ``in'' and the ``out'' states by two different symbols,
even though they belong to one and the same space $\mathbf \Phi$.

\subsection{The time evolution of the free bras and kets}
\label{sec:tidesolsfreebk}

In Ref.~\cite{IJTP03}, we constructed the free bras and kets, but we did not
obtain their time evolution. We do so here.

Either by direct calculation, or by making the potential zero in the 
time evolution of the Lippmann-Schwinger bras and kets, one 
can easily see that the free bras and kets evolve in time in the expected way:
\begin{eqnarray}
   \langle E|\, \rme ^{-\rmi H_0t/\hbar} = \rme ^{\rmi Et/\hbar} \langle E| 
           \, , \quad -\infty < t < \infty  \, ,   \\ [1ex]
         \rme ^{-\rmi H_0t/\hbar}|E\rangle = 
           \rme ^{-\rmi Et/\hbar} |E\rangle \, , 
           \quad -\infty < t < \infty  \, ,
\end{eqnarray}
where the free bras and kets act as the following integral operators (see 
Ref.~\cite{IJTP03}):
\begin{eqnarray}
      \langle E|\varphi  \rangle = \int_0^{\infty} \rmd r \,
        \varphi (r) \overline{\chi _0(r;E)} \, ,  \\ [1ex]
      \langle \varphi |E \rangle = \int_0^{\infty} \rmd r \,
        \overline{\varphi (r)} \chi _0(r;E) \, . 
\end{eqnarray}

\section{Other results of scattering theory}
\label{sec:otherreusst}

The ``in'' bras and kets can be used to expand the ``in'' states 
$\varphi ^+$. This expansion is the restriction of the eigenfunction 
expansion~(\ref{LSinvdiagonaliza}) to the space ${\mathbf \Phi}\equiv \Sw$, 
\begin{equation}
      \langle r|\varphi ^+\rangle =\int_0^{\infty}\rmd E\, 
      \langle r|E^+ \rangle \pebra \varphi ^+\rangle     \, .
        \label{8ibasivecph+}
\end{equation}
Similarly, by restricting Eq.~(\ref{LSinteexpre}) to 
${\mathbf \Phi}\equiv \Sw$, we obtain
\begin{equation}
     \pebra \varphi ^+\rangle =\int_0^{\infty}\rmd r\, 
      \pebra r\rangle \langle r|\varphi ^+\rangle       \, .
\end{equation}
The corresponding expansions of the $\psi ^-$ by the ``out'' bras and kets
follow  from the restriction of the eigenfunction 
expansions~(\ref{-LSinvdiagonaliza}) and (\ref{-LSinteexpre}) to $\Sw$:
\begin{eqnarray}
      &&\langle r|\psi ^-\rangle =\int_0^{\infty}\rmd E\, 
      \langle r|E^- \rangle \langle ^-E|\psi ^-\rangle    \, ,
              \label{8ibasivecph-}  \\
    &&\langle ^-E|\psi ^-\rangle =\int_0^{\infty}\rmd r\, 
       \langle ^-E|r\rangle  \langle r|\psi ^-\rangle  \, .
\end{eqnarray}
Expansions~(\ref{8ibasivecph+}) and (\ref{8ibasivecph-}) are the way through
which the RHS gives meaning to the formal expansions~(\ref{ibasivecph+}) and 
(\ref{ibasivecph-}).

The $S$-matrix element~(\ref{iprobaout+-}) can be written in terms of the 
action of the Lippmann-Schwinger bras and kets as in 
Eq.~(\ref{psi-ph+interofS}). The expansion~(\ref{psi-ph+interofS}) plays an 
important role in resonance theory and is proved in~\ref{sec:A3}. By a 
similar argument to that used to prove Eq.~(\ref{psi-ph+interofS}), one can
also prove Eq.~(\ref{starpbaisco0II}). Many formal identities follow from 
Eqs.~(\ref{psi-ph+interofS}) and (\ref{starpbaisco0II}). For instance,
\begin{equation}
      \langle E|S|E'\rangle = \langle ^-E|E{'}{^+} \rangle =
      S(E)\delta (E-E') \, ,
\end{equation}
\begin{equation}
      \int_0^{\infty}\rmd r \, \langle E|r\rangle \langle r|E'\rangle = 
      \int_0^{\infty}\rmd r \, 
       \langle ^{\pm}E|r\rangle \langle r|E'{^{\pm}}\rangle = 
      \delta (E-E') \, ,
\end{equation}
\begin{equation}
      \int_0^{\infty}\rmd r \, 
      \langle {^-}E|r\rangle \langle r|E'{^+} \rangle = 
       S(E) \delta (E-E') \, .
\end{equation}
As always, these expressions are to be understood within the RHS setting as 
part of a ``sandwich'' with well-behaved wave functions.

The Lippmann-Schwinger equations~(\ref{LSeq1}) and (\ref{LSeqbraspm}) are also 
understood as ``sandwiched'' with elements of $\mathbf \Phi$. For example,
Eq.~(\ref{LSeq1}) should be understood as
\begin{equation}
     \langle \varphi |E ^{\pm}\rangle = \langle \varphi |E\rangle +
     \langle \varphi |\frac{1}{E-H_0\pm \rmi \varepsilon}V|E^{\pm}\rangle \, ,
       \label{LSeq1-sand1}
\end{equation}
that is,
\begin{equation}
     \hskip-2.5cm
     \int_0^{\infty}\rmd r \, \overline{\varphi (r)} \chi ^{\pm}(r;E) =
    \int_0^{\infty}\rmd r \, \overline{\varphi (r)} \chi _0(r;E) +
 \int_0^{\infty}\rmd r \rmd s \, \overline{\varphi (r)} G_0^{\pm}(r,s;E) 
         V(s) \chi ^{\pm}(s;E) ,
       \label{LSeq1-sand2}
\end{equation}
where $\varphi$ is an element of $\mathbf \Phi$ that can be attached a 
superscript $+$ or $-$ depending on whether it is an ``in'' or an ``out''
wave function. Note that since both
$|E^+\rangle$ and $|E^-\rangle$ act on $\mathbf \Phi$, and since 
$|E\rangle$ is also well defined on $\mathbf \Phi$, the concerns raised
in~\cite{GADELLA-ORDONEZ} do not appear here. The concerns 
of~\cite{GADELLA-ORDONEZ} appear because in~\cite{GADELLA-ORDONEZ} it is
assumed that the Lippmann-Schwinger kets are functionals over two distinct 
spaces of Hardy functions, whereas in this paper the ``in'' and ``out'' kets 
both act on one and the same space $\mathbf \Phi$.

The M{\o}ller operators $\Omega _{\pm}$ can be expressed in terms of the 
operators $U_{\pm}$ and $U_0$ of Sec.~\ref{sec:DIdecomop} as~\cite{AMREIN}
\begin{equation}
      \Omega _{\pm}=U_{\pm}^{\dagger}U_0 \, .
     \label{omegaUs}
\end{equation}
As is well known, and as can be checked directly by using Eq.~(\ref{omegaUs}),
the M{\o}ller operators intertwine the total and the free Hamiltonians:
\begin{equation}
       H_0=\Omega _{\pm}^{\dagger}H\Omega _{\pm} \, .
\end{equation} 
Since our potential does not bind bound states, the M{\o}ller operators
are unitary operators on $L^2([0,\infty ),\rmd r)$. The well-known expression 
for the $S$-matrix operator in terms of the M{\o}ller operators then reads as
\begin{equation}
      S=\Omega _-^{\dagger} \Omega _+ =U_0^{\dagger}U_- U_{+}^{\dagger}U_0 \, .
      \label{smainmolerII}
\end{equation}
The operator $S$ is also a unitary operator on $L^2([0,\infty ),\rmd r)$. In 
the energy representation, the operator~(\ref{smainmolerII}) acts as 
multiplication by the function $S(E)={\mathcal J}_-(E)/{\mathcal J}_+(E)$. To 
be more precise, if we define the operator $\widehat{S}$ as
\begin{equation}
\begin{array}{rcl}
    \widehat{S}:L^2([0,\infty ),\rmd E) &\longmapsto & L^2([0,\infty ),\rmd E)
        \\
         \widehat{f}  &\longmapsto & 
      (\widehat{S}\widehat{f})(E)=S(E)\, \widehat{f}(E) \, ,
\end{array}
\end{equation}
then it can be proved (see~\ref{sec:A3}) that
\begin{equation}
       \widehat{S}=U_0SU_0^{-1} \, .
       \label{enerreSmaII}
\end{equation}

The M{\o}ller operators can be used to construct
the space ${\mathbf \Phi}_{0}$ of asymptotic ``in'' $\varphi ^{\rm in}$ and 
``out'' $\psi ^{\rm out}$ states,
\begin{equation}
      {\mathbf \Phi}_{\rm 0}=
      \Omega _{\pm}^{\dagger}{\mathbf \Phi} \, .
\end{equation}
A vector $\varphi ^{\rm in}$ belongs to ${\mathbf \Phi}_{0}$ if
\begin{equation}
      \pebra \varphi ^+ \rangle =\langle E|\varphi ^{\rm in} \rangle \, , 
\end{equation}
where $\varphi ^+ =\Omega _{+}\varphi ^{\rm in}$. A vector 
$\psi ^{\rm out}$ belongs to ${\mathbf \Phi}_{0}$ if
\begin{equation}
      \langle ^-E|\psi ^- \rangle =\langle E|\psi ^{\rm out} \rangle \, ,
\end{equation}
where $\psi ^- =\Omega _{-}\psi ^{\rm out}$. From the last two equations, it
follows that
\begin{equation}
      \Omega _{\pm}|E\rangle =|E^{\pm}\rangle \, ,
         \label{omegaaeeep}
\end{equation}
\begin{equation}
      \Omega_{\pm}=\int_0^{\infty}\rmd E\, |E^{\pm}\rangle \langle E| \, .
           \label{decomposeiosk}
\end{equation}
Again, Eqs.~(\ref{omegaaeeep}) and (\ref{decomposeiosk}) are to be understood 
as part of a ``sandwich'' with elements of ${\mathbf \Phi}$ and 
${\mathbf \Phi}_0$.

\section{Conclusions}
\label{sec:conclusions}

We have presented the RHS approach to the Lippmann-Schwinger 
equation. We have shown that the Lippmann-Schwinger bras are linear 
functionals that 
belong to the dual space ${\mathbf \Phi}^{\prime}$, whereas the
Lippmann-Schwinger kets are antilinear functionals that belong to the 
antidual space ${\mathbf \Phi}^{\times}$. To every 
Lippmann-Schwinger ket there corresponds a Lippmann-Schwinger bra, and vice 
versa. The Lippmann-Schwinger bras (kets) are left (right) eigenvectors of the 
Hamiltonian, and their time evolution is defined for all times. 

The following diagram summarizes the results concerning the ``in'' states
and kets:

\vskip0.3cm

\begin{equation}
    \hskip-2.7cm  \begin{array}{ccclccclccll}
    H_0; & \varphi ^{\rm in}(r) & \ & {\mathbf \Phi}_{0} & \subset & 
      L^2([0,\infty),\rmd r) &
      \subset & {\mathbf \Phi}_0^{\times} & \  & |E\rangle & \ &
      {\rm position \ repr.} \nonumber \\  [2ex]
       & & \ & \downarrow \Omega _+ &  &\downarrow \Omega _+  &
       & \downarrow \Omega _+^{\times} & \ & & &  \nonumber \\ [2ex]    
      H; & \varphi ^{+}(r) & \ & {\mathbf \Phi}& \subset & 
      L^2([0,\infty),\rmd r) &
      \subset & {\mathbf \Phi}^{\times} & \  & |E ^+\rangle & \ &
      {\rm position \ repr.} \nonumber \\  [2ex]
       & & \ & \downarrow U_+ &  &\downarrow U_+  &
       & \downarrow U_+^{\times} & \ & & &  \nonumber \\ [2ex]  
      \widehat{H}; & \widehat{\varphi}^+(E) & \ & 
      \widehat{\mathbf \Phi}_+ & 
       \subset & 
      L^2([0,\infty),\rmd E) & \subset & 
      \widehat{\mathbf \Phi}_+^{\times} & 
      \ & |\widehat{E}^+\rangle &\ & {\rm energy \ repr.}  \\ 
      \end{array}
\end{equation}

\vskip0.3cm

\noindent The results concerning the ``out'' states and kets are summarized by the following diagram:

\vskip0.3cm

\begin{equation}
     \hskip-2.7cm  \begin{array}{ccclccclccll}
      H_0; & \psi ^{\rm out}(r) & \ & {\mathbf \Phi}_{0} & \subset & 
      L^2([0,\infty),\rmd r) &
      \subset & {\mathbf \Phi}_{0}^{\times} & \  & |E\rangle & \ &
      {\rm position \ repr.} \nonumber \\  [2ex]
       & & \ & \downarrow \Omega _- &  &\downarrow \Omega _-  &
       & \downarrow \Omega _-^{\times} & \ & & &  \nonumber \\ [2ex]    
      H; & \psi ^{-}(r) & \ & {\mathbf \Phi} & \subset & 
      L^2([0,\infty),\rmd r) &
      \subset & {\mathbf \Phi}^{\times} & \  &  |E ^-\rangle & \ &
      {\rm position \ repr.} \nonumber \\  [2ex]
       & & \ & \downarrow U_- &  &\downarrow U_-  &
       & \downarrow U_-^{\times} & \ & & &  \nonumber \\ [2ex]  
      \widehat{H}; & \widehat{\psi}^-(E) & \ & 
      \widehat{\mathbf \Phi}_- & 
       \subset & 
      L^2([0,\infty),\rmd E) & \subset & 
      \widehat{\mathbf \Phi}_-^{\times} & 
      \ &  |\widehat{E}^-\rangle &\ & {\rm energy \ repr.}  \\ 
      \end{array}
\end{equation}

\vskip0.3cm

\noindent Analogous diagrams summarize the results for the ``in'' and for
the ``out'' bras.

For the sake of simplicity, we have proved our results within the example of 
the spherical shell potential and for zero angular momentum. Nonetheless, 
with obvious modifications, the results of this paper remain valid
for higher partial waves and for a large class of spherically symmetric
potentials that includes, in particular, potentials of finite range. Therefore,
we conclude that the natural mathematical setting for the solutions of the 
Lippmann-Schwinger equation is the rigged Hilbert space rather than just the 
Hilbert space. That rigged Hilbert space, however, seems to be unrelated
to the rigged Hilbert spaces of Hardy functions 
of~\cite{GADELLA84,BG,GADELLA-ORDONEZ,BRTK,CAGADA,GAGO,DIS}.

\ack

It is a great pleasure to acknowledge many fruitful conversations with
Alfonso Mondrag\'on over the past several years. Additional discussions
with J.~G.~Muga, M.~Gadella, A.~Bohm, I.~Egusquiza, R.~de la Llave, L.~Vega and
R.~Escobedo are also acknowledged.

This research was supported by MEC fellowship No.~SD2004-0003.


{
\appendix

\section{The Sturm-Liouville theory}
\label{sec:A1}

In this paper, we have used several theorems that form the backbone of the
Sturm-Liouville theory. These theorems can be found in the treatise of
Dunford and Schwartz~\cite{DUNFORD}. For the sake of completeness, we recall
those theorems in this appendix.

The following theorem provides the procedure to obtain the Green 
function of $H$ (cf.~Theorem~XIII.3.16 of Ref.~\cite{DUNFORD}):

\vskip0.5cm

\newtheorem*{Theo1}{Theorem~1}
\begin{Theo1} \quad Let $H$ be the self-adjoint operator (\ref{LSoperator}) 
derived from the real formal differential operator (\ref{doh}) by the 
imposition of the boundary condition (\ref{sac}). Let 
${\rm Im}(E) \neq 0$. Then there is exactly one solution $\chi (r;E)$ of 
$(h-E)\sigma =0$ square-integrable at $0$ and satisfying the boundary 
condition (\ref{sac}), and exactly one solution $f(r;E)$ of $(h-E)\sigma =0$ 
square-integrable at infinity. The resolvent $(E-H)^{-1}$ is an integral 
operator whose kernel $G(r,s;E)$ is given by
\begin{equation}
       G(r,s;E)=\left\{ \begin{array}{ll}
               \frac{2m}{\hbar ^2} \,
      \frac{\chi (r;E) \, f(s;E)}{W(\chi ,f)}
               &r<s      \\ [1ex]
      \frac{2m}{\hbar ^2} \,
      \frac{\chi (s;E) \, f(r;E)}{W(\chi ,f)}
                       &r>s  \, ,
                  \end{array} 
                 \right. 
	\label{exofGFA}
\end{equation}
where $W(\chi ,f)$ is the Wronskian of $\chi$ and $f$
\begin{equation}
       W(\chi ,f)=\chi f '-\chi ' f\, .
\end{equation}
\end{Theo1}

\vskip0.5cm

The following theorem provides the operators $U_{\pm}$ 
(cf.~Theorem XIII.5.13 of Ref.~\cite{DUNFORD}):

\vskip0.5cm

\newtheorem*{Theo2}{Theorem~2}
\begin{Theo2}(Weyl-Kodaira) \quad Let $h$ be the formally self-adjoint 
differential operator (\ref{doh}) defined on the interval 
$[0,\infty )$. Let $H$ be the self-adjoint operator (\ref{LSoperator}). Let
$\Lambda$ be an open interval of the real axis, and suppose that there 
is given a set $\left\{ \sigma _1(r;E),\, \sigma _2(r;E)\right\}$ of 
functions, defined 
and continuous on $(0,\infty )\times \Lambda$, such that for each fixed 
$E$ in $\Lambda$, $\left\{ \sigma _1(r;E),\, \sigma _2(r;E)\right\}$ forms 
a basis for
the space of solutions of $h\sigma =E\sigma$. Then there exists a 
positive $2\times 2$ matrix measure $\left\{ \rho _{ij} \right\}$ defined on
$\Lambda$, such that
\begin{enumerate}
      \item the limit 
     \begin{equation}
      (U f)_i(E)=\lim_{c\to 0}\lim_{d\to \infty} 
        \left[ \int_c^d f(r) \overline{\sigma _i(r;E)}\rmd r \right]
      \end{equation}
     exists in the topology of $L^2(\Lambda ,\left\{ \rho _{ij}\right\})$ 
    for each 
    $f$ in $L^2([0,\infty ),\rmd r)$ and defines an isometric isomorphism $U$ 
    of ${\sf E}(\Lambda )L^2([0,\infty ),\rmd r)$ onto 
    $L^2(\Lambda ,\left\{ \rho _{ij} \right\})$, 
    where ${\sf E}(\Lambda )$ is the 
    spectral projection associated with $\Lambda$;
      \item for each Borel function $G$ defined on the real line and vanishing
       outside $\Lambda$,
    \begin{equation}
      U{\cal D}(G(H))=\left\{ [f_i]\in 
       L^2(\Lambda ,\left\{ \rho _{ij}\right\}) \, | \
           [Gf_i]\in L^2(\Lambda ,\left\{ \rho _{ij}\right\}) \right\}
     \end{equation}
    and
    \begin{equation}
      \hskip-1.5cm (UG(H)f)_i(E)=G(E)(Uf)_i(E), \quad i=1,2, \, E\in \Lambda ,
       \, f\in {\cal D}(G(H)) \, .
            \label{Gfunction}
    \end{equation}  
\end{enumerate}
\end{Theo2}

\vskip0.5cm

The following theorem provides the inverses of $U_{\pm}$ 
(cf.~Theorem XIII.5.14 of Ref.~\cite{DUNFORD}):

\vskip0.5cm

\newtheorem*{Theo3}{Theorem~3}
\begin{Theo3}(Weyl-Kodaira)\quad  Let $H$, $\Lambda$, 
$\left\{ \rho _{ij} \right\}$, etc., be as in Theorem~2. Let $E_0$ and 
$E_1$ be the end points of $\Lambda$. Then
\begin{enumerate}
      \item the inverse of the isometric isomorphism $U$ of 
      $E(\Lambda )L^2([0,\infty ),\rmd r)$ onto 
      $L^2(\Lambda ,\left\{ \rho _{ij}\right\})$ is 
      given by the formula
     \begin{equation}
       (U^{-1}F)(r)=\lim_{\mu _0 \to E_0}\lim_{\mu _1 \to E_1}
       \int_{\mu _0}^{\mu _1} \left( \sum_{i,j=1}^{2}
             F_i(E)\sigma _j(r;E)\rho _{ij}(\rmd E) \right)
        \label{inverseU}
     \end{equation}
    where $F=[F_1,F_2]\in L^2(\Lambda ,\left\{ \rho _{ij}\right\})$, the 
   limit existing
   in the topology of $L^2([0, \infty ),\rmd r)$;
    \item if $G$ is a bounded Borel function vanishing outside a Borel set 
    $e$ whose closure is compact and contained in $\Lambda$, then $G(H)$ has 
    the representation
   \begin{equation}
       G(H)f(r)=\int _0^{\infty}f(s)K(H,r,s)\rmd s \, , 
   \end{equation}
   where
   \begin{equation}
      K(H,r,s)=\sum_{i,j=1}^2 \int_e 
       G(E)\overline{\sigma _i(s;E)}\sigma _j(r;E)\rho _{ij}(\rmd E) \, .
   \end{equation}
\end{enumerate}
\end{Theo3}

\vskip0.5cm

The spectral measures are provided by the following theorem 
(cf.~Theorem XIII.5.18 of Ref.~\cite{DUNFORD}):

\vskip0.5cm

\newtheorem*{Theo4}{Theorem~4}
\begin{Theo4}(Titchmarsh-Kodaira)\quad  Let $\Lambda$ be an open interval 
of the real axis and $O$ be an open set in the complex plane containing 
$\Lambda$. Let $\left\{ \sigma _1(r;E),\, \sigma _2(r;E)\right\}$ be a set 
of functions 
which form a basis for the solutions of the equation $h\sigma =E\sigma$, 
$E\in O$, and which are continuous on $(0,\infty )\times O$ and analytically 
dependent on $E$ for $E$ in $O$. Suppose that the kernel $G(r,s;E)$ for the 
resolvent $(E-H)^{-1}$ has a representation
\begin{equation}
      G(r,s;E)=\left\{ \begin{array}{lll}
                   \sum_{i,j=1}^2 \theta _{ij}^-(E)\sigma _i(r;E)
                   \overline{\sigma _j(s;\overline{E})}\, , &
                     \qquad & r<s \, ,  \\  [1ex]
                  \sum_{i,j=1}^2 \theta _{ij}^+(E)\sigma _i(r;E)
                   \overline{\sigma _j(s;\overline{E})} \, ,&\qquad & r>s \, ,
                  \end{array}
                 \right.
       \label{Gitofsigma}
\end{equation}
for all $E$ in ${\rm Re}(H)\cap O$, and that $\left\{ \rho _{ij} \right\}$ is 
a positive matrix measure on $\Lambda$ associated with $H$ as in 
Theorem~2. Then
the functions $\theta _{ij}^{\pm}$ are analytic in ${\rm Re}(H)\cap O$, and
given any bounded open interval $(E_1,E_2)\subset \Lambda$, we have for 
$1\leq i,j\leq 2$,
\begin{equation}
     \hskip-1.5cm   \begin{array}{lll}
       \rho _{ij}((E_1,E_2))&=& \lim_{\delta \to 0}\lim_{\epsilon \to 0+}
         \frac{1}{2\pi \rmi}\int_{E_1+\delta}^{E_2-\delta}
          [ \theta _{ij}^-(E-\rmi \epsilon )-\theta _{ij}^-(E+\rmi \epsilon )
          ]\rmd E \\ [1ex]
      \quad &=& \lim_{\delta \to 0}\lim_{\epsilon \to 0+}
         \frac{1}{2\pi \rmi}\int_{E_1+\delta}^{E_2-\delta}
          [ \theta _{ij}^+(E-\rmi \epsilon )-\theta _{ij}^+(E+\rmi \epsilon )
          ]\rmd E \, .
         \end{array} 
       \label{measures}
\end{equation}
\end{Theo4}

\section{List of auxiliary functions}
\label{sec:A2}

The coefficients in Eq.~(\ref{LSchi}) are given by
\begin{eqnarray}
  & \hskip-1cm & {\cal J}_1(E)= {\cal J}_1(k) =\frac{1}{2}\rme^{-\rmi\kappa a} 
     \left( \sin (ka)+
      \frac{k}{\rmi\kappa}
      \cos (ka) \right) , \label{Jfunctions1} \\
 & \hskip-1cm &{\cal J} _2 (E) = {\cal J}_2(k)=\frac{1}{2}\rme^{\rmi\kappa a} 
     \left( \sin (ka)-
      \frac{k}{\rmi\kappa}
      \cos (ka) \right) , \\
    & \hskip-1cm &   {\cal J}_3(E)= {\cal J}_3(k)=\frac{1}{2}\rme^{-\rmi kb}
                   \left[
            \left(1+ \frac{\kappa}{k} \right) 
       \rme^{\rmi \kappa b}{\cal J}_1(k)+
       \left(1-\frac{\kappa}{k} \right) 
       \rme^{-\rmi\kappa b}{\cal J}_2(k) 
        \right] , \\
     & \hskip-1cm & {\cal J}_4(E)= {\cal J}_4(k)=\frac{1}{2}\rme^{\rmi kb}
                   \left[
            \left(1-\frac{\kappa}{k} \right) 
       \rme^{\rmi \kappa b}{\cal J} _1(k)+
       \left(1+\frac{\kappa}{k}\right) 
        \rme^{-\rmi\kappa b}{\cal J}_2(k) 
        \right] ,  
        \label{Jfunctions4}
\end{eqnarray}
where $k$ is given by Eq.~(\ref{wavenumber}) and $\kappa$ is given by 
\begin{equation}
      \kappa = \sqrt{\frac{2m}{\hbar ^2} (E-V_0) \,} \, .
\end{equation}
The coefficients in Eq.~(\ref{fpmfun}) are given by
\begin{eqnarray}
     & \hskip-1cm & {\cal A}_3^{\pm}(E)={\cal A}_3^{\pm}(k)=
           \frac{1}{2}\rme^{-\rmi\kappa b} 
     \left(1\pm \frac{k}{\kappa}\right) \rme^{\pm \rmi kb} ,
        \label{Afunctions3} \\
     & \hskip-1cm & {\cal A}_4^{\pm}(E)={\cal A}_4^{\pm}(k)=
            \frac{1}{2}\rme^{\rmi\kappa b}
       \left(1\mp \frac{k}{\kappa} \right)
       \rme^{\pm \rmi kb} , \\
      & \hskip-1cm & {\cal A}_1^{\pm}(E)={\cal A}_1^{\pm}(k)=
          \frac{1}{2}\rme^{-\rmi ka}
                   \left[
      \left(1+\frac{\kappa}{k}\right) 
      \rme^{\rmi\kappa a}{\cal A}_3^{\pm}(k)+
     \left(1-\frac{\kappa}{k} \right) 
      \rme^{-\rmi\kappa a} {\cal A}_4^{\pm}(k)
     \right] , \\ 
     & \hskip-1cm & {\cal A}_2^{\pm}(E)={\cal A}_2^{\pm}(k)=
            \frac{1}{2}\rme^{\rmi ka}
                   \left[
     \left(1-\frac{\kappa}{k} \right) 
     \rme^{\rmi\kappa a}{\cal A}_3^{\pm}(k)+
     \left(1+\frac{\kappa}{k} \right) 
      \rme^{-\rmi\kappa a}{\cal A}_4^{\pm}(k)
     \right] .
         \label{Afunctions2} 
\end{eqnarray}
The coefficients in Eq.~(\ref{LSsigam2cos}) are given by
\begin{eqnarray}
      & \hskip-1cm & {\cal C}_1(E)= {\cal C}_1(k)=
       \frac{1}{2}\rme^{-\rmi\kappa a}
             \left( \cos (ka)-\frac{k}{\rmi\kappa} \sin (ka) \right), 
        \label{Cfunctions1} \\
      & \hskip-1cm & {\cal C}_2(E)= {\cal C}_2(k)=
       \frac{1}{2}\rme^{\rmi\kappa a}
        \left( \cos (ka)+\frac{k}{\rmi\kappa} \sin (ka) \right), \\
      & \hskip-1cm & {\cal C}_3(E)= {\cal C}_3(k)=\frac{1}{2}\rme^{-\rmi kb}
                   \left[
      \left(1+\frac{\kappa}{k}\right) \rme^{\rmi\kappa b}{\cal C}_1(k)+
     \left(1-\frac{\kappa}{k} \right) \rme^{-\rmi\kappa b}{\cal C}_2(k)
     \right] , \\ 
     & \hskip-1cm & {\cal C}_4(E)= {\cal C}_4(k)=\frac{1}{2}\rme^{\rmi kb}
                   \left[
     \left(1-\frac{\kappa}{k} \right) \rme^{\rmi\kappa b}{\cal C}_1(k)+
     \left(1+\frac{\kappa}{k} \right) 
      \rme^{-\rmi\kappa b}{\cal C}_2(k)
     \right]  . 
      \label{Cfunctions4}   
\end{eqnarray}

\section{Proofs}
\label{sec:A3}

Here we prove some results we invoked throughout the paper. In
the proofs, it will be convenient to denote the dual and the antidual 
extensions of $H$ ($U_+$) by, respectively, $H^{\times}$ ($U_+^{\times}$) and 
$H^{\prime}$ ($U_+^{\prime}$), in order to make clear that the operator $H$ 
($U_+$) is acting outside the Hilbert space.

\vskip0.5cm

\begin{proof}[Proof of Proposition~1] ({\it i}),
({\it ii}) and ({\it iii}) were proved in Refs.~\cite{DIS,FP02}.

\vskip0.3cm

({\it iv}) From definition~(\ref{LSdefinitionket+E}), it is easy 
to see that $|E ^+\rangle$ is an antilinear functional. In order to show that 
$|E^+\rangle$ is $\mathbf \Phi$-continuous, we define
\begin{equation}
      {\cal C}^+(E):= \sup _{r\in [0,\infty )} \left| \chi ^+(r;E) \right| \, .
\end{equation}
From the expression for $\chi ^+(r;E)$ in Eq.~(\ref{pmeigndu}), it is clear 
that ${\cal C}^+(E)$ is a finite number for each energy $E$. Now, because
\begin{eqnarray}
      |\langle \varphi ^+ |E^+\rangle | &=&
       \left| \int_0^{\infty}\rmd r \, \overline{\varphi ^+(r)}
       \chi ^+(r;E)\right| \nonumber \\
      &\leq & \int_0^{\infty}\rmd r \, |\overline{\varphi ^+(r)}| 
         |\chi ^+(r;E)| \nonumber \\
      &\leq & {\cal C}^+(E) \int _0^{\infty}\rmd r \, |\varphi ^+(r)| 
           \nonumber \\
      &=& {\cal C}^+(E) \int_0^{\infty}\rmd r \,
      \frac{1}{1+r} (1+r) |\varphi ^+(r)| \nonumber \\
      &\leq & {\cal C}^+(E) \left( \int_0^{\infty}\rmd r \, 
      \frac{1}{(1+r)^2} \right) ^{1/2} 
      \left( \int_0^{\infty}\rmd r \, 
      \left| (1+r) \varphi ^+(r) \right| ^2 \right) ^{1/2} \nonumber \\
       &=&{\cal C}^+(E) \| \varphi ^+ \| _{1,0} \, ,
      \label{contikeEp}
\end{eqnarray}
the functional $|E^+\rangle$ is $\mathbf \Phi$-continuous. In a similar
way, one can prove that $|E^-\rangle$ is also a $\mathbf \Phi$-continuous 
antilinear functional.

It is clear from definition~(\ref{defibraEplus}) that $\pebra$ is a linear 
functional over $\Sw$. Because
\begin{eqnarray}
      \left| \pebra \varphi ^+ \rangle \right| &=
      \left| \langle \varphi ^+ |E^{+}\rangle \right|  
       \hskip1cm & \mbox{by (\ref{baccka})}   \nonumber \\
      &\leq  {\cal C}^+(E) \| \varphi ^+ \| _{1,0} \, ,
           \hskip1cm & \mbox{by (\ref{contikeEp})}
\end{eqnarray}
the functional $\pebra$ is $\mathbf \Phi$-continuous. That
$\langle ^-E|$ is also a $\mathbf \Phi$-continuous linear functional can
be proved in a similar way.

\vskip0.3cm

({\it v}) In order to prove that $|E^+\rangle$ is a (generalized) eigenvector 
of $H$, we make use of the conditions satisfied by the elements of 
$\Sw$ at $r=0,\infty$:
\begin{eqnarray}
       \langle \varphi ^+|H^{\times}|E^+\rangle &=& 
          \langle H \varphi ^+|E^+\rangle \nonumber \\
       &=& \int_0^{\infty}\rmd r \, 
       \left( -\frac{\hbar ^2}{2m}\frac{\rmd ^2}{\rmd r^2}+V(r) \right)
       \overline{\varphi ^+(r)} \chi ^+(r;E) \nonumber \\
       &=&-\frac{\hbar ^2}{2m}
       \left[ \frac{\rmd \overline{\varphi ^+(r)}}{\rmd r} \chi ^+(r;E) 
       \right] _0^{\infty} 
       +\frac{\hbar ^2}{2m}
       \left[ \overline{\varphi ^+(r)} \frac{\rmd \chi ^+(r;E)}{\rmd r} 
       \right] _0^{\infty} \nonumber \\ 
       &&+ \int_0^{\infty}\rmd r \, \overline{\varphi ^+(r)}
       \left( -\frac{\hbar ^2}{2m}\frac{\rmd ^2}{\rmd r^2}+V(r) \right)
         \chi ^+(r;E)
        \nonumber \\
       &=&\int_0^{\infty}\rmd r \, \overline{\varphi ^+(r)}
       \left( -\frac{\hbar ^2}{2m}\frac{\rmd ^2}{\rmd r^2}+V(r) \right)
         \chi ^+(r;E)
        \nonumber \\
      &=&E\langle \varphi ^+ |E^+\rangle \, ,
         \label{eigeneEpk}
\end{eqnarray}
where we have used conditions~(\ref{mainisexi}) and (\ref{zeroinflimit}) in
the next to the last step and Eq.~(\ref{schroeque}) in the last step. The
proof that $|E^-\rangle$ is an eigenvector of $H$ follows the pattern
of Eq.~(\ref{eigeneEpk}).

On the other hand, because
\begin{eqnarray}
     \pebra H^{\prime}|\varphi ^+ \rangle &=
     \pebra H\varphi ^+ \rangle \nonumber \\
       &= \overline{\langle H\varphi ^+| E^{+}\rangle} 
         \hskip1cm & \mbox{by (\ref{baccka})}    \nonumber \\
      &= E \, \overline{\langle \varphi ^+| E^{+}\rangle}
           \hskip1cm & \mbox{by (\ref{eigeneEpk})}        \nonumber \\
      &= E \pebra \varphi ^+ \rangle  \, ,
          \hskip1cm & \mbox{by (\ref{baccka})}
       \label{brapeiH}
\end{eqnarray}
the bra $\pebra$ is a (generalized) left eigenvector of $H$. An argument
similar to that in Eq.~(\ref{brapeiH}) shows that $\langle ^-E|$ is also a 
left eigenvector of $H$.

\vskip0.3cm

({\it vi}) Because
\begin{eqnarray}
       \langle \widehat{\varphi}^+|\widehat{E}^+\rangle 
    &= \langle \widehat{\varphi}^+|U_+^{\times}|E^+\rangle  \nonumber \\
     &= \langle U_+^{\dagger}\widehat{\varphi}^+|E^+\rangle  \nonumber \\ 
     &= \langle \varphi^+|E^+\rangle  \nonumber \\ 
     &= \overline{\widehat{\varphi}^+(E)}  \, , 
\end{eqnarray}
$|\widehat{E}^+\rangle$ acts as the antilinear Schwartz delta functional. A
similar argument shows that $|\widehat{E}^-\rangle$ also acts as the 
antilinear Schwartz delta functional.

Equation~(\ref{pluslSdf}) follows from
\begin{equation}
   \langle ^+\widehat{E}|\widehat{\varphi}^+ \rangle =
    \overline{\langle \widehat{\varphi}^+ |\widehat{E}^{+}\rangle}=  
         \widehat{\varphi}^+(E) \, .
     \label{cahinks}
\end{equation}
Finally, Eq.~(\ref{minuslSdf}) follows from a chain of equalities similar
to (\ref{cahinks}).
\end{proof}

\vskip0.5cm

\begin{proof}[Proof of Eq.~(\ref{enerreSmaII})] In order to prove 
Eq.~(\ref{enerreSmaII}), we first prove that 
\begin{equation}
       (U_-g)(E)=S(E)(U_+g)(E) \, , \quad  g\in L^2([0,\infty ),\rmd r) \, .
      \label{prelimsursoII}
\end{equation}
Since by Eq.~(\ref{chi+tSe-})
\begin{equation}
       \chi ^+(r;E)=S(E) \chi ^-(r;E)\, ,
\end{equation}
and since
\begin{equation}
       \overline{S(E)}=\frac{1}{S(E)} \, , \quad E>0 \, ,
\end{equation}
we conclude that
\begin{equation}
       \chi ^-(r;E)=\overline{S(E)} \chi ^+(r;E)\, .
      \label{minusovlplusII}
\end{equation}
By substituting Eq.~(\ref{minusovlplusII}) into the integral 
expression~(\ref{-LSinteexpre}) for the operator $U_-$, we get to
\begin{equation}
       \hskip-1cm
       (U_-g)(E)=\int_0^{\infty}\rmd r \, g(r)S(E)\overline{\chi ^+(r;E)} =
       S(E) \int_0^{\infty}\rmd r \, g(r) \overline{\chi ^+(r;E)}  \, .
      \label{equainlinsiuII}
\end{equation}
Comparison of (\ref{equainlinsiuII}) with (\ref{LSinteexpre}) leads to 
(\ref{prelimsursoII}).

Now,
\begin{eqnarray}
      (U_0SU_0^{-1})\widehat{f}&=
          (U_0U_0^{\dagger}U_-U_+^{\dagger}U_0U_0^{-1})\widehat{f} 
             & \hskip1cm \mbox{by (\ref{smainmolerII})}
                 \nonumber \\
      &= (U_-U_+^{\dagger})\widehat{f} 
          & \quad  \nonumber \\
      &= S(E) (U_+U_+^{\dagger})\widehat{f} 
               &  \hskip1cm  \mbox{by (\ref{prelimsursoII})}   \nonumber \\
      &=\widehat{S}\widehat{f} \, ,  & \quad 
\end{eqnarray}
which proves~(\ref{enerreSmaII}).
\end{proof}

\vskip0.5cm

\begin{proof}[Proof of Eq.~(\ref{psi-ph+interofS})] Let
$\psi ^- ,\varphi ^+ \in {\mathbf \Phi}$. Since
$\psi ^-$ and $\varphi ^+$ belong, in particular, to 
$L^2([0,\infty ),\rmd r)$, we can let the unitary operator $U_-$ act on both 
of them,
\begin{equation}
      (\psi ^-,\varphi ^+)=(U_-\psi ^-,U_-\varphi ^+) \, .
\end{equation}
The vectors $U_-\psi ^-$ and $U_-\varphi ^+$ belong to 
$L^2([0,\infty ),\rmd E)$. Therefore,
\begin{equation}
     (\psi ^-,\varphi ^+)= (U_-\psi ^-,U_-\varphi ^+)=\int_0^{\infty}\rmd E \,
     \overline{(U_-\psi ^-)(E)} (U_-\varphi ^+)(E) \, .
\end{equation}
From Eq.~(\ref{prelimsursoII}) it follows that
\begin{equation}
      (U_-\varphi ^+)(E)=S(E)(U_+\varphi ^+)(E) \, .
\end{equation}
Thus,
\begin{equation}
      (\psi ^-,\varphi ^+)=\int_0^{\infty}\rmd E \,
     \overline{(U_-\psi ^-)(E)}S(E) (U_+\varphi ^+)(E) \, .
     \label{indremkso1II}
\end{equation}
Since $\psi ^- , \varphi ^+  \in {\mathbf \Phi}$, we are allowed to write
\begin{equation}
      \overline{(U_-\psi ^-)(E)}=\langle \psi ^-|E^- \rangle \, .
     \label{indremkso2II}
\end{equation}
\begin{equation}
      (U_+\varphi ^+)(E)=\pebra \varphi ^+\rangle \, .
      \label{indremkso3II}
\end{equation}
Substitution of (\ref{indremkso2II}) and (\ref{indremkso3II}) into 
(\ref{indremkso1II}) leads to (\ref{psi-ph+interofS}).
\end{proof}

}


\vskip1cm

\end{document}